\documentclass{aa}

\usepackage{graphicx}
\usepackage{txfonts}
\usepackage{csquotes}
\begin{document} 

   \title{X-ray radiative transfer in full 3D with \sc{skirt}}

   \author{Bert Vander Meulen
          \inst{1}
          \and
          Peter Camps
          \inst{1}
          \and
          Marko Stalevski
          \inst{1}
          \inst{2}
          \and
          Maarten Baes
          \inst{1}
          }
   \institute{Sterrenkundig Observatorium, Universiteit Gent, Krijgslaan 281 S9, 9000 Gent, Belgium\\
              \email{bert.vandermeulen@ugent.be}
              \and
              Astronomical Observatory, Volgina 7, 11060 Belgrade, Serbia
        }

   \date{Received December 23, 2022; accepted April 19, 2023}

 
  \abstract
   {Models of active galactic nuclei (AGN) suggest that their circumnuclear media are complex with clumps and filaments, while recent observations hint towards polar extended structures of gas and dust, as opposed to the classical torus paradigm. The X-ray band could form an interesting observational window to study these circumnuclear media in great detail.
   }
   {We want to extend the radiative transfer code \textsc{skirt} with the X-ray processes that govern the broadband X-ray spectra of obscured AGN, to study the structure of AGN circumnuclear media in full 3D, based on their reflected X-ray emission.
   }
   {We extend the \textsc{skirt} code with Compton scattering on free electrons, photo-absorption and fluorescence by cold atomic gas, scattering on bound electrons, and extinction by dust. This includes a novel treatment of extreme-forward scattering by dust, and a detailed description of anomalous Rayleigh scattering. To verify our X-ray implementation, we perform the first dedicated benchmark of X-ray torus models, comparing five X-ray radiative transfer codes.
   }
   {The resulting radiative transfer code covers the X-ray to mm wavelength range self-consistently, has all the features of the established \textsc{skirt} framework, is publicly available, and is fully optimised to operate in arbitrary 3D geometries. In the X-ray regime, we find an excellent agreement with the simulation results of the \textsc{MYTorus} and \textsc{RefleX} codes, which validates our X-ray implementation. We find some discrepancies with other codes, which illustrates the complexity of X-ray radiative transfer and motivates the need for a robust framework that can handle non-linear 3D radiative transfer effects. We illustrate the 3D nature of the code by producing synthetic X-ray images and spectra of clumpy torus models.}
   {\textsc{skirt} forms a powerful new tool to model circumnuclear media in 3D, and make predictions for the X-ray band in addition to the dust-dominated IR-to-UV range. The new X-ray functionalities of the \textsc{skirt} code allow for uncomplicated access to a broad suite of 3D X-ray models for AGN, that can easily be tested and modified. This will be particularly useful with the advent of X-ray microcalorimeter observations in the near future. 
   }


   \maketitle
   
%

\section{Introduction}
\label{sect:intro}
Large amounts of gas and dust can be found in the central regions of most active galaxies. These ambient media play a crucial role as they provide the accretion reservoir powering active galactic nuclei (AGN), and they reprocess the optical, ultraviolet (UV), and X-ray emission of the central engine. Through these circumnuclear media, the large-scale AGN feedback on the host galaxy can be studied, and the conditions close to the supermassive black hole can be probed \citep{ramosalmeida17}. Yet, the detailed characteristics of these regions and the physics behind them remain poorly understood. According to the AGN unification scheme, dusty gas exists as a toroidal structure in the equatorial plane, which explains the observed variety of AGN types as a distribution of system orientations relative to the observer \citep{antonucci93, urry95, netzer15}. This putative dusty torus is assumed to absorb 40 to 70 percent of the accretion disk emission, which is re-emitted at thermal infrared (IR) wavelengths peaking at about $10$ to $20~\text{$\mu$m}$ \citep[see e.g.][and references therein]{netzer15, stalevski16}. The successful modelling of observational IR spectra using dusty torus models supports this unified picture of obscuration in local Seyfert-type AGN \citep[][]{siebenmorgen05, hao05, spoon07, schweitzer08, mendozacastrejon15, hatziminaoglou15}.

However, recent high angular resolution mid-IR observations of AGN suggest that the distribution of obscuring dust is also extended in the polar direction, as opposed to the classical dusty torus paradigm \citep{honig12, lopezgonzaga16, asmus16}. These polar dust features are now firmly established in two nearby active galaxies, NGC 1068 \citep[][]{jaffe04, wittkowski04, raban09, lopezgonzaga14} and Circinus Galaxy \citep[][]{tristram07, tristram14, stalevski19, isbell22} while similar mid-IR extensions have been confirmed for other AGN \citep[][]{honig13, asmus19, leftley19}. The physical origin of these polar structures could be radiation-driven winds that are launched close to the dust sublimation radius \citep{honig12}, as demonstrated in theoretical calculations by \citet[][]{gallagher15, chan16, wada16, vollmer18}. Furthermore, detailed radiative transfer modelling suggests that the AGN-obscuring medium has a complex structure with clumps and filaments, possibly forming an intricate two-phase medium of clump and inter-clump regions \citep{stalevski12, stalevski16}.

A promising way to study gas and dust in active galaxies is to look at their X-ray emission, as the interactions between X-rays and the circumnuclear medium produce characteristic spectral features that contain information on this obscuring body of gas and dust \citep{ricci14, ichigawa19}. Observations reveal strong X-ray emission emerging from most AGN, which is likely produced in a corona of hot electrons close to the central black hole, where optical and UV disk photons are Compton up-scattered to X-ray energies \citep[][]{haardt91}. This X-ray component is detected even for heavily obscured sources, which opens an interesting observational window on AGN obscuration \citep[][]{ricci17}. While the transmitted X-ray continuum displays absorption features linked to the intervening gas density and abundances, most information is encoded in the reflected X-ray emission, which is now used to constrain the putative torus geometry and its physical properties \citep[][]{gupta21, yamada21, osorioclavijo22}. The main signature of X-ray reflection in obscured AGN is the appearance of various fluorescent emission lines, the most prominent one being the narrow \element[ ][]{Fe}~K$_\alpha$ line at $6.4~\text{keV}$ \citep[][]{mushotzky78, nandra89, pounds89, fukazawa11}. In addition, X-ray scattering produces characteristic spectral features such as a Compton reflection hump peaking at about $30~\text{keV}$, and a Compton shoulder at the low energy side of bright fluorescent lines, which are believed to form powerful probes on the circumnuclear material in AGN \citep[][]{matt02, matt03, yaqoob11, odaka16, Hikitani18}.

In order to fully exploit the diagnostic power of these reprocessed X-ray features, one has to model the relevant X-ray absorption, scattering, and re-emission processes in great detail. Furthermore, to make realistic model predictions for AGN circumnuclear media, one needs to perform radiative transfer simulations in three dimensions. The detailed three-dimensional substructure of the obscuring medium has been shown to have a strong effect on the dust modelling results in the IR-to-UV wavelength range, which was investigated with radiative transfer simulations based on smooth torus models \citep{fritz06}, clumpy torus models \citep{nenkova08a, nenkova08b, honig06, schartmann08}, two-phase filamentary torus models \citep{stalevski12}, and most recently two-component disk-plus-wind models \citep{honig17, stalevski17, stalevski19, honig19}. Yet, the X-ray signatures of these three-dimensional torus models have not been studied to the same level of sophistication as it has been done in the infrared. The introduction of advanced X-ray torus models consistent with the 3D IR dust models would allow for a joint analysis of the X-ray and IR bands, which could break some degeneracies that are encountered when dissecting dusty structures using infrared observations only \citep[][]{farrah16, ogawa19, tanimoto19, esparzaarredondo21}. Finally, the advent of X-ray microcalorimeter observations with \emph{XRISM}/Resolve \citep[][]{tashiro20} and \emph{Athena}/X-IFU \citep[][]{barret18} could revolutionise our understanding of AGN obscuration, given that detailed X-ray models are available that can be used to leverage the high energy resolution to constrain their precise distribution of gas and dust.

However, reprocessed X-ray spectra are often fitted with one-dimensional \textsc{xspec} models \citep{arnaud96} such as \textsc{pexrav} \citep{magdziarz95} and \textsc{pexmon} \citep{nandra07}, which cannot capture the three-dimensional complexity of the circumnuclear medium. More advanced X-ray spectral models are available \citep[][]{murphy09, ikeda09, brightman11, odaka11, odaka16, liu14, furui16, paltani17, balokovic18, balokovic19, tanimoto19, buchner19, buchner21}, but either they have a limited geometrical flexibility, or they do not incorporate the full suite of X-ray physics. To our knowledge, there is no radiative transfer code that combines advanced X-ray radiative processes such as bound-electron scattering and interactions with dust grains, with a full-3D treatment of radiative transfer that overcomes geometrical limitations such as the need for 3D transfer media to exhibit certain spatial symmetries, to follow a 2D distribution of spherical clumps, or to be assembled as a superposition of 1D and 2D building-block components.

To this end, we extend the 3D radiative transfer code \textsc{skirt} \citep{camps20} with the X-ray physical processes in gas and dust, to obtain an X-ray radiative transfer code with all benefits of the established \textsc{skirt} framework. Beyond the study of circumnuclear media, these X-ray extensions allow for various potential applications of the \textsc{skirt} code in X-ray astrophysical research, such as dust mineralogy with \emph{XRISM} \citep[][]{corrales16, xrism20}, modelling clumpy nova ejecta in supersoft X-ray sources \citep{ness22}, studying gas and dust in the interstellar medium \citep[][]{costantini19}, constraining AGN contributions in galaxy SED fitting \citep{yang20, yang22}, and more. 

This work presents the new X-ray functionalities of the \textsc{skirt} radiative transfer code, with a particular focus on gas and dust in active galactic nuclei. The outline for this paper is as follows: In Sect.~\ref{sect:MCRT}, we introduce the \textsc{skirt} radiative transfer code. In Sect.~\ref{sect:processes}, we present the new X-ray processes and their implementation to the \textsc{skirt} code. In Sect.~\ref{sect:benchmark}, we verify our X-ray implementation by comparing \textsc{skirt} with some well-established spectral models. Finally, we demonstrate the 3D capabilities of \textsc{skirt} in the X-ray regime by calculating images and spectra of clumpy torus models in Sect.~\ref{sect:demonstration}. In Sect.~\ref{sect:discussion}, we summarise our results and discuss their implications.

\section{Monte Carlo radiative transfer with \sc{skirt}}
\label{sect:MCRT}
Monte Carlo radiative transfer (MCRT) is a direct simulation method to study the effects of absorption, scattering, and re-emission in complex transfer media. In this method, the radiation field is represented as a flow of discrete photon packets, and all relevant physics are emulated by drawing random events from the appropriate probability distributions \citep{MCRT1, MCRT3}. The MCRT technique is the most popular numerical method to study radiative transfer in general 3D geometries, partly due to the existence of various acceleration mechanisms that are based on importance sampling \citep{MCRT2}. One of these optimisations is the photon peel-off technique to detect flux contributions, which enables efficient simulations in full 3D \citep{peeloff}.

\textsc{skirt} is a state-of-the-art Monte Carlo radiative transfer code, developed and maintained at Ghent University \citep{baes03, baes11, camps15, camps20}. Historically, the \textsc{skirt} code has focused on dusty astrophysical systems, modelling scattering, absorption, and re-emission by dust. More recently, the \textsc{skirt} code was extended to support radiative transfer in electron and gas media, in addition to dust \citep{camps20}. \textsc{skirt} offers a large suite of built-in geometries \citep{baes15}, radiation sources, and medium characterisations, in addition to interfaces for post-processing hydrodynamical simulations based on the SPH, AMR, or Voronoi paradigms \citep{HSPP1, HSPP2, popping22}. Furthermore, hierarchical octree and k-d~tree algorithms are available for efficiently gridding complex transfer geometries \citep{grid1, grid2}. \textsc{skirt} has been successfully applied in different fields of astrophysics, with recent applications in galaxies \citep[e.g.][]{popping22, shen22, vijayan22, degraaf22, camps22, trcka22, jang22} and AGN \citep[e.g.][]{stalevski19, AGN2, AGN3, isbell22, stalevski22}. The \textsc{skirt} C++ code is open-source, well-documented\footnote{\url{https://skirt.ugent.be}}, and publicly available online\footnote{\url{https://github.com/SKIRT/SKIRT9}}, with tutorials for both users and developers.

The latest major code redesign (\textsc{skirt} version 9) introduced polarisation by aligned dust grains and line transfer \citep{camps20, polarised, line}. Furthermore, this \textsc{skirt} version supports advanced kinematics, modelling the effect of bulk velocities and velocity dispersions, for both sources and transfer media. In Sect.~\ref{sect:processes}, new X-ray processes will be described in the local rest frame of the transfer medium, while \textsc{skirt} tracks the flow of photons in the model rest frame (i.e.\ the model coordinate system). This implies that photon energies are transformed to the moving medium frame before each interaction, and back to the model frame after the interaction:
\begin{equation}
    E' = E \, \left(1 - \frac{\vec{k} \cdot \vec{\upsilon}}{c}\right),
    \label{eq:doppler}
\end{equation}
with $\vec{\upsilon}$ the medium velocity, $\vec{k}$ the photon propagation direction, and $E$ the photon energy in the model frame. In this way, kinematic effects will be self-consistently incorporated into the \textsc{skirt} radiative transfer results.

\textsc{skirt} implements a large number of state-of-the-art acceleration mechanisms to speed up its MCRT calculations. Most notably, the photon peel-off technique \citep{peeloff} allows for recording flux contributions at each scattering event, as opposed to collecting photon packets as they escape the simulated model. In addition to realising a significant speed up, this peel-off technique also avoids the spurious blurring otherwise caused by averaging over a finite solid angle around the observer direction. Other advanced techniques include path length stretching, forced scattering, continuous absorption, composite biasing and explicit absorption \citep{MCRT2, baes16, baes22}. Combined, these mechanisms drastically speed up the \textsc{skirt} simulations (as compared to other radiative transfer codes), especially in complex 3D geometries. Moreover, the \textsc{skirt} code supports a hybrid parallelisation scheme, combining multi-threading and multi-processing, which allows for efficient simulations on systems ranging from laptops to high-performance computing facilities \citep{parallel}.

Up to this point, the \textsc{skirt} code was operational in the mm to far-UV wavelength range. In this work, we extend the advanced treatment of radiative transfer into the X-ray range, obtaining an X-ray MCRT code with all benefits of the established \textsc{skirt} framework. Furthermore, this allows for radiative transfer simulations that self-consistently cover the X-ray band and the dust-dominated IR-to-UV range, linking X-ray reprocessing to dust modelling in AGN obscuring media.

\section{X-ray radiative processes}
\label{sect:processes}
Focusing on the radiative processes that govern the broadband spectra of obscured AGN in the $0.1$ to $500~\text{keV}$ range, we implement Compton scattering on free electrons, photo-absorption and fluorescence by cold atomic gas, scattering on bound electrons, and extinction by dust. This includes a detailed description of anomalous Rayleigh scattering (Sect.~\ref{sect:BES}) and extreme-forward scattering by dust (Sect.~\ref{sect:extremeforwardscattering}), which goes beyond the X-ray physics that is currently implemented in AGN obscuration models available for spectral fitting.
\subsection{Free-electron media}
\label{sect:CS}
Compton scattering describes the inelastic scattering of high-energy photons on free electrons, which forms the relativistic extension of Thomson scattering into the X-ray domain. Compton scattering has a differential scattering cross section given by the \citet{Klein-Nishina} formula, which is a function of the incoming photon energy:
\begin{equation}
    \frac{d\sigma_{\text{KN}}}{d\Omega}(\theta, x)=\frac{3\, \sigma_\text{T}}{16\pi}\left[C^3(\theta, x) + C(\theta, x) -C^2(\theta, x)\sin^2\theta\right], \label{differentialkleinnishina}
\end{equation}
with $x = E/m_ec^2$ the photon energy scaled to the electron rest energy, $\sigma_\text{T} \approx 6.65 \times 10^{-25}~\text{cm}^2$ the total Thomson cross section, and $C(\theta, x)$ the Compton factor equal to:
\begin{equation}
    C(\theta, x) = {\Big({1+x \, \left(1-\cos \theta\right)}\Big).}^{-1} \label{comptonfactor}
\end{equation}

Integration of the differential cross section yields the total Compton scattering cross section as a function of the incoming photon energy:
\begin{equation}
    \sigma_\text{KN}(x) = \frac{3\, \sigma_\text{T}}{4} \left[\frac{1+x}{(1+2x)^2} + \frac{2}{x^2} + \left(\frac{1}{2x}-\frac{x+1}{x^3}\right) \ln(2x+1) \right].
    \label{totalcompton}
\end{equation}
Compared to Thomson scattering, the Compton cross section is reduced at higher energies. However, Compton scattering will still form the dominant source of X-ray extinction above $12~\text{keV}$, see Sect.~\ref{sect:totalgasextinction}. By normalising the Klein-Nishina formula Eq.~(\ref{differentialkleinnishina}), we obtain the Compton scattering phase function:
\begin{equation}
        \Phi(\theta, x) = \frac{1}{\sigma_\text{KN}(x)}\frac{d\sigma_\text{KN}}{d\Omega}\left(\theta, x\right). \label{Comptonphasefunction}
\end{equation}
The energy dependence of this phase function is significant above $10~\text{keV}$ and should not be neglected in detailed radiative transfer simulations. Eq.~(\ref{Comptonphasefunction}) predicts a bias towards forward scattering, that increases with increasing photon energy. Even then, Compton scattering is not restricted to forward directions only, and the exact phase function shape should be implemented.

Unlike Thomson scattering, photon energies will change by Compton scattering on free electrons. Conserving the four-momentum of the photon-electron pair, one has:
\begin{equation}
    E_\text{out} = C(\theta, x)\; E_\text{in},
    \label{eq:ComptonEchange}
\end{equation}
with $C(\theta, x)$ the Compton factor as given by Eq.~(\ref{comptonfactor}). In the electron rest frame, photons can only lose energy upon interacting with a free electron. The relative energy loss will be most significant for high-energy photons, and increases with increasing scattering angle. Forward scattering does not change the photon energy, while backscattering maximally reduces the energy by a factor of $1+2x$. In result, the observed X-ray spectrum at a certain photon energy will be formed by photon interactions at higher energies, which must be included in the radiative transfer simulations to obtain accurate reflection spectra, see Sect.~\ref{sect:totalgasextinction}. Below $0.25~\text{keV}$, the energy shift becomes negligible ($<0.1\%$) and Compton scattering converges to elastic Thomson scattering, with a differential cross section:
\begin{equation}
    \frac{d\sigma_\text{T}}{d\Omega}(\theta)=\frac{3\, \sigma_\text{T}}{16\pi}\Big[1 + \cos^2\theta\Big], \label{DifferentialThomson}
\end{equation}
which does not depend on the incoming photon energy.

In \textsc{skirt}, the spatial distribution of free electrons is described by an electron number density $n_e(\Vec{r})$, which defines the transfer geometry of the free-electron medium. We introduce Compton scattering to the \textsc{skirt} code using the exact formulae presented in this section. The scattering phase function (Eq.~(\ref{Comptonphasefunction})) is an intricate function of the photon energy, which cannot be inverted analytically to generate random scattering angles \citep{devroye}. However, this phase function can be sampled efficiently by using a variation on Khan's technique, as described by \citet{hua}. This algorithm combines the composition and rejection methods to avoid computationally expensive operations, with a rejection rate of about $30\%$ depending on the photon energy.

\subsection{Cold-gas media}
\subsubsection{Photo-absorption}
\label{sect:PA}
In this work, photo-absorption describes the absorption of X-ray photons in cold atomic gas, where the photon energy is used to liberate a bound electron from a neutral gas atom. Photo-absorption forms the main source of X-ray extinction below $12~\text{keV}$ (see Sect.~\ref{sect:totalgasextinction}), but drops in importance at higher energies. This causes Compton-thick AGN to have most of their soft X-ray emission obscured up to some energy, which depends on the total amount of intervening gas.

The photo-absorption cross section $\sigma_Z(E)$ of a given atom $Z$ is just the sum of its electronic subshell contributions:
\begin{equation}
    \sigma_{Z}(E) = \sum_{nl} \: \sigma_{Z,\, nl}(E), \label{eq:PA_one_element}
\end{equation}
which displays spectral features reflecting the internal electron structure. Photo-absorption cross sections $\sigma_{Z,\, nl}(E)$ for individual electronic subshells $nl$ are presented by \citet{vern1}, who provide analytic fits\footnote{\url{https://www.pa.uky.edu/~verner/photo.html}} to the cross sections of all ground-state subshells of the first 30 neutral elements (\element[][]{H} to \element[][]{Zn}), based on Hartree-Dirac-Slater calculations. Updated cross sections for the outer shells of 18 selected elements\footnote{\element[][]{H} to \element[][]{Si}, \element[][]{S}, \element[][]{Ar}, \element[][]{Ca}, and \element[][]{Fe} as included in \citet{OP}.} are provided by \citet{vern2} based on R-matrix calculations, which are valid up to some maximum energy $E_{\text{max}}$. This update is most important for neutral \element[][]{He}, having just one electron K-shell, which dominates the neutral gas extinction in the $0.1$ to $0.5 \, \text{keV}$ range. For neutral hydrogen, both references list the exact same cross section. \textsc{skirt} incorporates the \citet{vern1} subshell cross sections for all 30 elements, updated with the available \citet{vern2} outer-shell data for energies up to $E_{\text{max}}$. This implementation is identical to the \textsc{phabs} model in \textsc{xspec} \citep{arnaud96}, although \textsc{skirt} does include 12 more elements.\footnote{\textsc{phabs} includes 18 elements: \element[][]{H}, \element[][]{He}, \element[][]{C}, \element[][]{N}, \element[][]{O}, \element[][]{Ne}, \element[][]{Na}, \element[][]{Mg}, \element[][]{Al}, \element[][]{Si}, \element[][]{S}, \element[][]{Cl}, \element[][]{Ar}, \element[][]{Ca}, \element[][]{Cr}, \element[][]{Fe}, \element[][]{Co}, and \element[][]{Ni}. The 12 elements not implemented in \textsc{phabs} contribute up to $1\%$ to the total photo-absorption cross section.}

Combined with the abundances $a_Z$ for each element, the total photo-absorption cross section per \element[][]{H}-atom is calculated as:
\begin{equation}
    \sigma_\text{PA}(E) = \sum_{Z} \: a_Z \, \left(1-\beta_Z\right) \, \left(\sum_{nl} \: \sigma_{Z,\, nl}(E)\right), \label{eq:PA}
\end{equation}
with $a_Z$ the number density of element $Z$ relative to \ion{H}{I}, and $1-\beta_Z$ the fraction of this element found in the gas phase (i.e.\ not locked in dust grains, see Sect.~\ref{sect:DU}). In \textsc{skirt}, \citet{angr} solar abundances $a_Z$ are used as the default atomic gas composition. In addition, we allow for custom gas mixes by considering the abundance table as an input parameter. Finally, the spatial distribution of the cold gas medium is characterised by a neutral hydrogen number density $n_\text{H}(\Vec{r})$.

\begin{figure}
\centering
\includegraphics[width=\hsize]{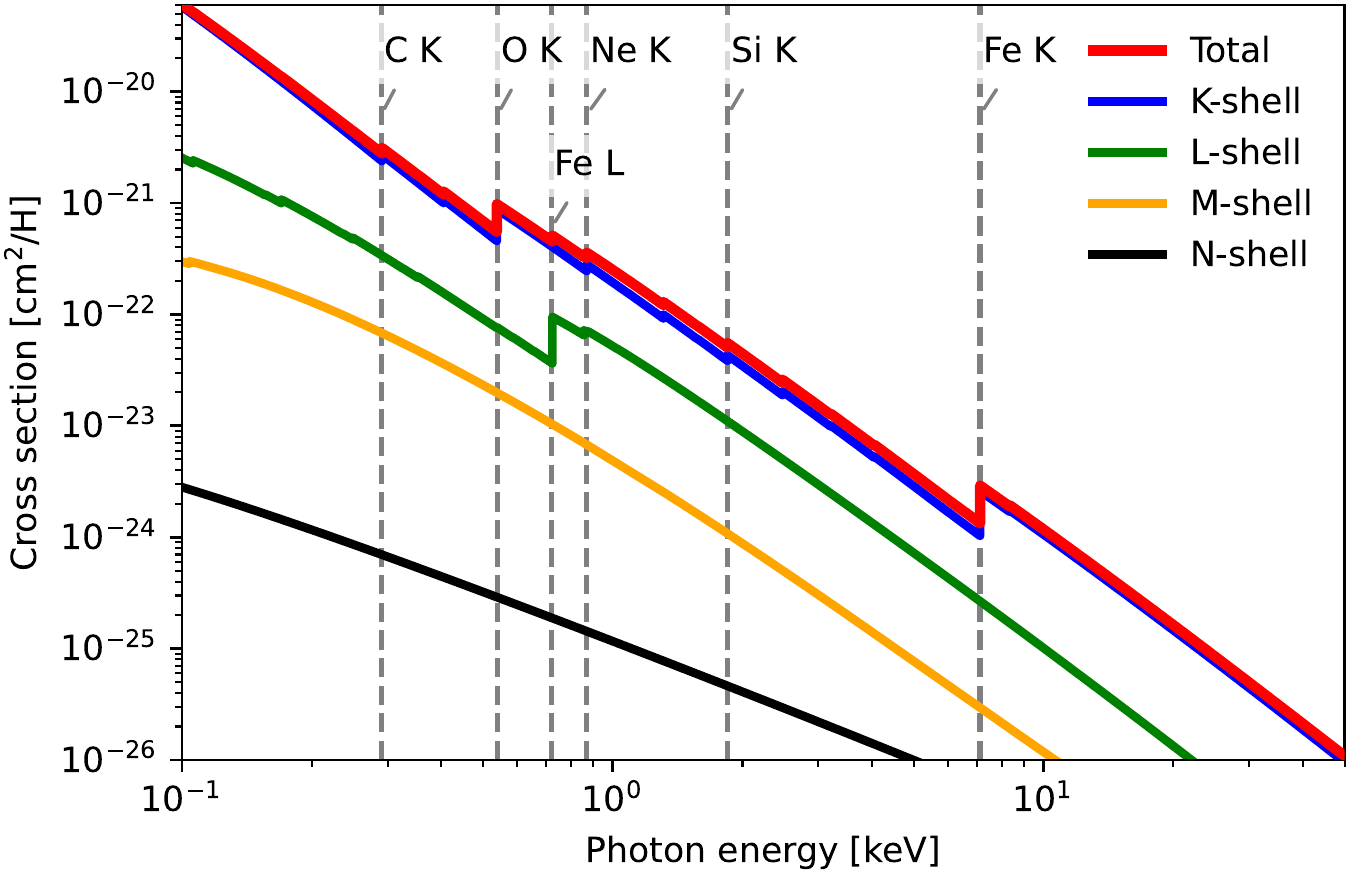}
  \caption{Total photo-absorption cross section for a cold gas medium with \citet{angr} abundances and no dust. The K- and L-shells dominate the absorption over the entire X-ray range. The prominent absorption edge structure reflects the atomic composition of the gas.}
    \label{fig:totalPA}
\end{figure}
The total photo-absorption cross section for a cold gas medium is shown in Fig.~\ref{fig:totalPA}, assuming \citet{angr} abundances and no dust depletion. This cross section shows characteristic absorption edges where the individual elements start to contribute to the total gas absorption. Most prominent are the \element[][]{O}~K-edge at $0.538~\text{keV}$ and the \element[ ][]{Fe}~K-edge at $7.124~\text{keV}$, while the \element[ ][]{Fe}~L-edge at $0.724~\text{keV}$ is the only significant L-shell feature.

In general, gas velocity dispersions must be accounted for by convolving all subshell cross sections $\sigma_{Z,\, nl}(E)$ with a Gaussian kernel of appropriate width.\footnote{In the atom frame, the photon energy is perceived as given by Eq.~(\ref{eq:doppler}), with $\vec{k} \cdot \vec{\upsilon}$ distributed as a Gaussian with a width $\upsilon_\text{th} = \sqrt{{k_\text{B}T}/{m}}$, which is the thermal velocity for a gas of temperature $T$ and particle mass $m$.} In practice however, these cross sections are smooth functions of energy, except for the discontinuity at the absorption edge. Therefore, the effect of a convolution is limited to energies close to the edge energy $E_\text{edge}$, smoothing the sharp step to a sigmoid. This can be incorporated into the subshell cross sections as:
\begin{equation}
    \sigma_{Z,\, nl}'(E) = \frac{\sigma_{Z,\, nl}(E_\text{edge} +2 E_\text{th})} {2} \, \left[ 1+
    \mathrm{erf} \left( \frac{E - E_\text{edge}} {E_\text{th}} \right) \right],
\end{equation}
for energies between $E_\text{edge} -2 E_\text{th}$ and $E_\text{edge} +2 E_\text{th}$, with $E_\text{th}$ the energy dispersion corresponding to the thermal velocity of the absorbing atom\footnote{$E_\text{th} = \left(\upsilon_\text{th}/c\right) E_\text{edge}$, with $\upsilon_\text{th}$ the thermal velocity.}, considering that the convolution of a step function with a Gaussian is given by the error function. This approximation to incorporate velocity dispersions achieves good accuracy and continuity around the edges of all important subshells. However, note how the neutral atomic material surrounding AGN is expected to have gas temperatures less than $10^{3.5}~\text{K}$ \citep[see e.g.][]{wada16}, which makes the corresponding smoothing unresolvable with current X-ray instrumentation. Nevertheless, the effect of these velocity dispersions can be studied in great detail with \textsc{skirt}, which will be especially valuable for studying ionised gas media at higher temperatures in the future.

Figure~\ref{fig:totalPA} also shows the contributions of photo-absorption by different electronic shells. In the X-ray range, the dominant absorption process is K-shell photo-absorption, where an inner-shell electron is ejected from the atom. These K-shell electrons have binding energies in the $0.1 - 10~\text{keV}$ range, where their photo-absorption cross sections peak. Absorption by higher-shell electrons kicks in at lower energies, and quickly becomes more unlikely with increasing energy: $\sigma_{Z,\, nl}(E) \propto E^{-3}$. While L-shell absorption is still important for some heavy elements (mainly \element[][]{Fe}), the total contribution of M- and N-shells is less than $3.5 \%$ over the entire range where photo-absorption is relevant (i.e.\ below $40~\text{keV}$, see Sect.~\ref{sect:totalgasextinction}). However, including these subshells does not slow down our radiative transfer calculations, as the total cross section for a configured gas mix is calculated only once during simulation setup.

\begin{figure}
\centering
\includegraphics[width=\hsize]{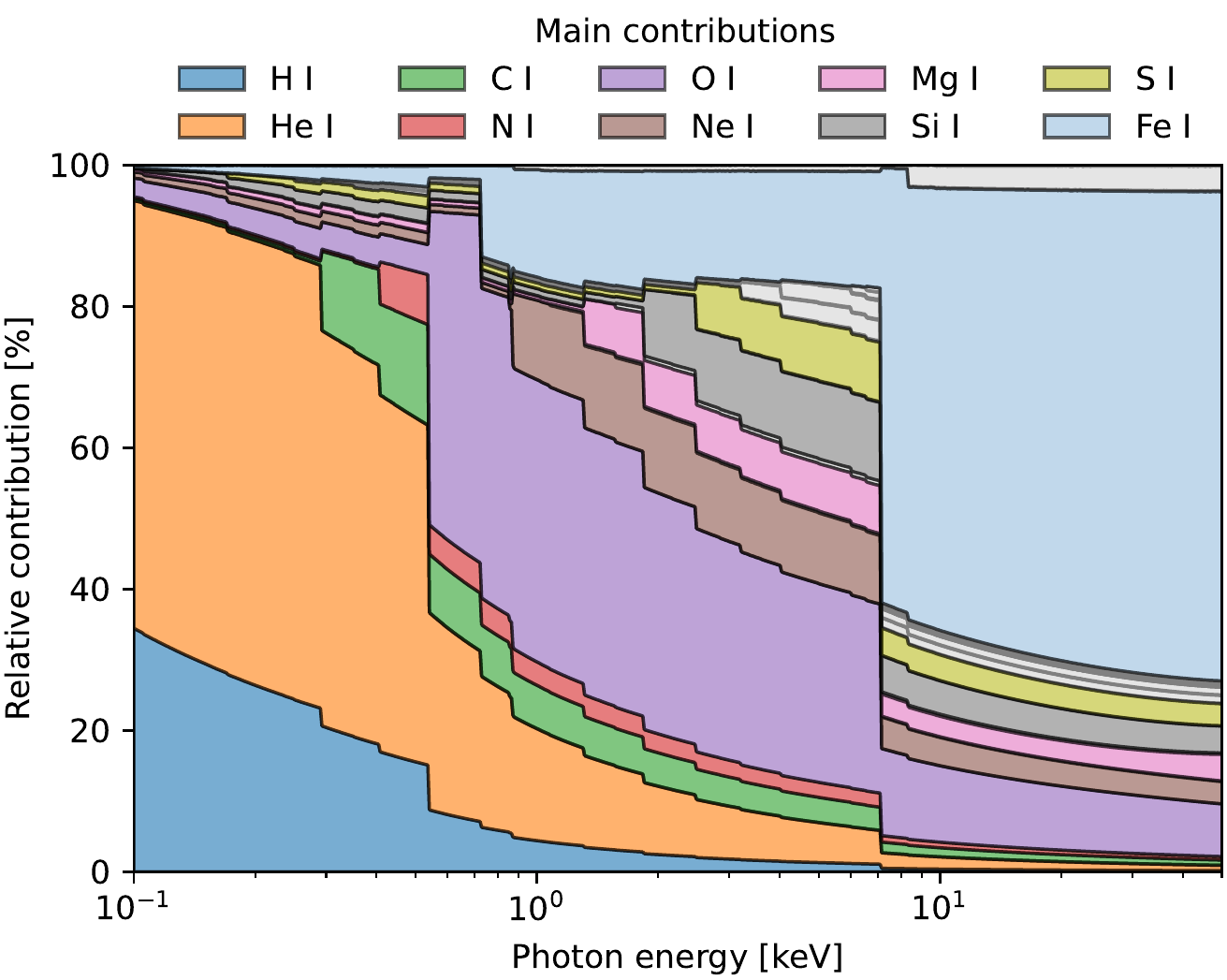}
  \caption{Main contributions to the total photo-absorption cross section of a cold gas (see Fig.~\ref{fig:totalPA}), assuming \citet{angr} abundances and no dust.}
     \label{fig:contributionsPA}
\end{figure}
Figure~\ref{fig:contributionsPA} shows the relative contributions of each element to the total photo-absorption cross section. Overall, the main contributing elements are just the elements that are most abundant. Below $0.5~\text{keV}$, the photo-absorption is dominated by helium, and therefore it was important that we incorporated the updated outer-shell cross section of \element[][]{He} as provided by \citet{vern2}. At higher energies, oxygen becomes the most important element, up to $7.1~\text{keV}$ where K-shell absorption by iron kicks in. In spite of its dominant abundance, the contribution of neutral hydrogen is marginal in the X-ray range above $1~\text{keV}$ ($< 5\%$).

Currently, \textsc{skirt} focuses on photo-absorption by neutral atomic gas, modelling the cold and distant material that causes most of the X-ray extinction in Compton-thick AGN. This same cold gas produces the characteristic X-ray reflection spectrum that is used to study heavily obscured sources, and is possibly located at the distant inner wall of the dusty torus (Sect~\ref{sect:intro}). In this work, we are not implementing radiative transfer in ionised plasmas, which can be found more closely to the central AGN. However, we verified how the photo-absorption cross sections of mildly-ionised material differ only slightly from the neutral cross sections, allowing \textsc{skirt} to be used in these types of media. Note how interested users can easily add ionised media to the open-source developing environment of the \textsc{skirt} code.\footnote{A developer guide is available on the \textsc{skirt} website: \url{https://skirt.ugent.be}} Finally, we did not consider neutral elements with $Z>30$, which can be neglected because of their much lower abundances.

\subsubsection{Fluorescence}
\label{sect:FL}
Fluorescence forms one possible way for gas atoms to relax after a photo-absorption event. After an inner-shell electron has been ejected by the absorption of an X-ray photon, a higher-shell electron can fall into this inner-shell vacancy, emitting the energy difference as a fluorescent photon. As fluorescence describes bound-bound transitions in atoms, characteristic line photons will be emitted, which produce fluorescent lines in the X-ray spectra of AGN. The most prominent fluorescent line is the \element[ ][]{Fe}~K$_\alpha$ line at $6.4~\text{keV}$, which forms a strong, characteristic feature in the reprocessed X-ray emission of Compton-thick AGN \citep{mushotzky78, nandra89, pounds89}. A competing, non-radiative relaxation channel is described by the Auger process, which involves the ejection of one or more electrons, with no effect on the radiation field.

\begin{table}
\caption{Fluorescent line transitions that are implemented in \textsc{skirt} for all 30 elements in the photo-absorbing gas, with their corresponding electronic transition. The associated fluorescent yields and line energies for \ion{Fe}{I} are listed as an illustration. Note how \citet{bearden67} lists equal line energies for the \element[][]{Fe} K$_{\beta1}$ and \element[][]{Fe} K$_{\beta3}$ sublines.}
\label{table:1}
\centering                                     
\begin{tabular}{c c c c}          
\hline\hline                       
Line & Transition & \ion{Fe}{I} yield $[1]$& \ion{Fe}{I} energy $[\text{keV}]$\\    
\hline                                  
    K$_{\alpha1}$ & L$_3 \to$ K & $Y_{\element[ ][]{Fe}, \, \mathrm{K}_{\alpha1}} = 0.199$ & $E_{\element[ ][]{Fe}, \, \mathrm{K}_{\alpha1}} = 6.40$\\     
    K$_{\alpha2}$ & L$_2 \to$ K & $Y_{\element[ ][]{Fe}, \, \mathrm{K}_{\alpha2}} = 0.101$ & $E_{\element[ ][]{Fe}, \, \mathrm{K}_{\alpha2}} = 6.39$\\
    K$_{\beta1}$ & M$_3 \to$ K & $Y_{\element[ ][]{Fe}, \, \mathrm{K}_{\beta1}} = 0.024$ & $E_{\element[ ][]{Fe}, \, \mathrm{K}_{\beta1}} = 7.06$\\
    K$_{\beta3}$ & M$_2 \to$ K & $Y_{\element[ ][]{Fe}, \, \mathrm{K}_{\beta3}} = 0.012$ & $E_{\element[ ][]{Fe}, \, \mathrm{K}_{\beta3}} = 7.06$\\
\hline                                             
\end{tabular}
\end{table}

In the X-ray range, we focus on K-shell fluorescence only, which is the radiative relaxation after a K-shell photo-absorption (see Sect.~\ref{sect:PA}). Table~\ref{table:1} lists the fluorescent transitions that are implemented in \textsc{skirt}, with their corresponding line identifiers. K$_\alpha$-transitions can occur in neutral atoms with $Z>5$, while K$_\beta$-transitions are possible for $Z>12$. The probability of a fluorescent transition $K_i$ towards a given K-shell vacancy is expressed by the fluorescent yield $Y_{Z,\,K_i}$, which is provided by \citet{EADL} for all elements in the photo-absorbing gas.\footnote{\url{https://www-nds.iaea.org/epics2014/ENDL/EADL.ALL}} As the L-shell photo-absorption cross sections and fluorescent yields are much lower than the corresponding K-shell values, L-shell fluorescence can be neglected in the X-ray range. Likewise, M- and N-shell fluorescence are negligible relative to the K-shell.

\begin{figure}
\centering
\includegraphics[width=\hsize]{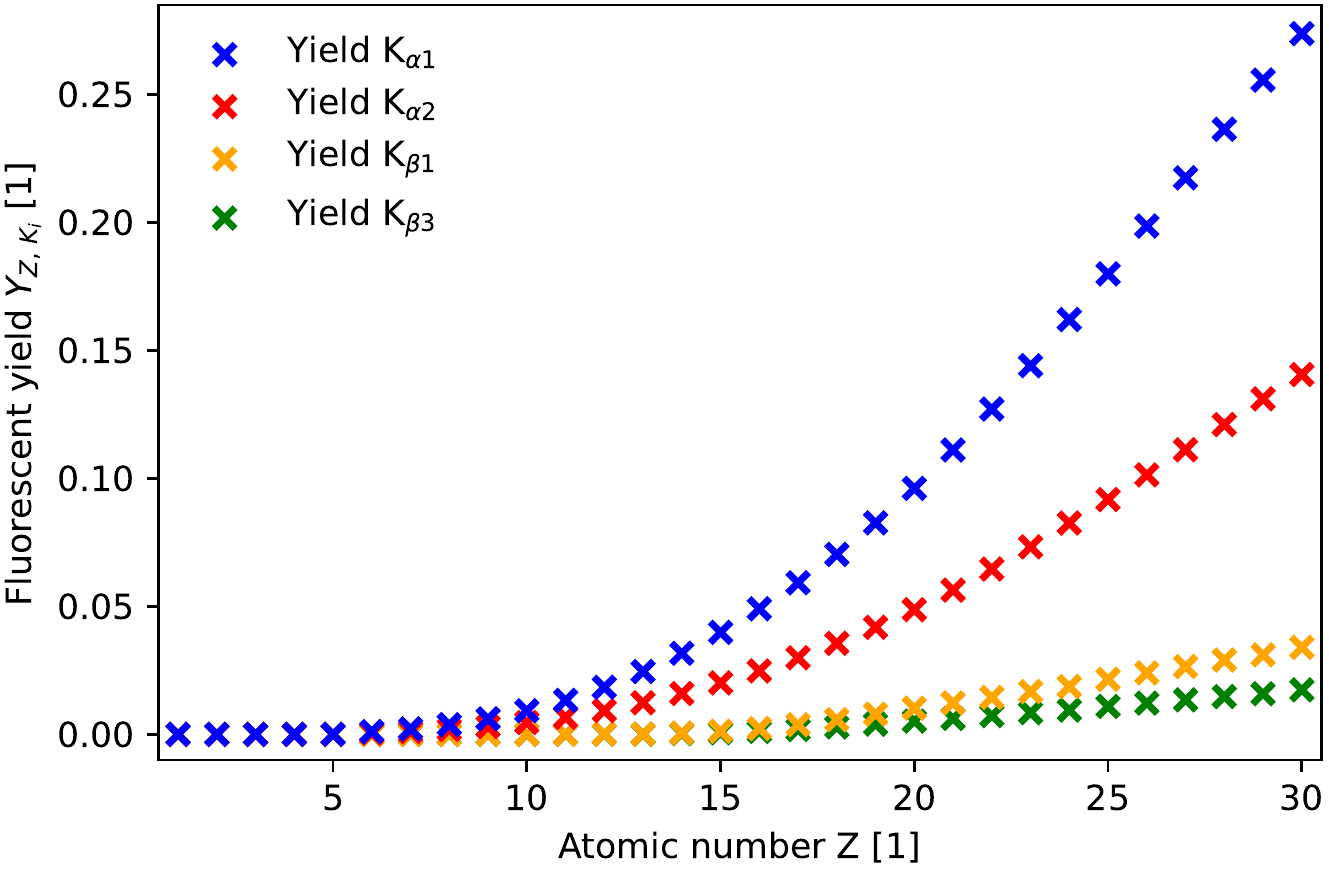}
  \caption{Fluorescent yields for all fluorescent transitions that are implemented in \textsc{skirt}, as provided by \citet{EADL}. See Table~\ref{table:1} for the transition definitions.}
     \label{fig:yields}
\end{figure}
Figure~\ref{fig:yields} shows the fluorescent yields for all transitions that are incorporated in \textsc{skirt}. For each element, the K$_{\alpha1}$-transition has the highest probability, producing a fluorescent line that is two times stronger than the corresponding K$_{\alpha2}$-line, and 5.5 times stronger than both K$_{\beta}$-lines combined. As the fluorescent yields increase along the atomic sequence, abundant heavy elements like \element[][]{Fe} and \element[][]{Ni} will dominate the fluorescent line emission of AGN. For each transition, the corresponding line energy is taken from \citet{bearden67}, who provides lab measurements that should be more accurate than the theoretical values as calculated by \citet{EADL}. As the K$_{\alpha1}$- and K$_{\alpha2}$-lines have very similar line energies, they are often observed as a single K$_{\alpha}$-line with current CCD instruments. For neutral iron, both sublines are separated by $13~\text{eV}$ (see Table~\ref{table:1}), which could be resolved with upcoming microcalorimeter observations \citep[][]{barret18, tashiro20}. Analogously, one observes a single K$_{\beta}$-line at slightly higher photon energies.

In principle, fluorescence is a re-emission process that could be performed in the simulation after calculating the total photo-absorption in each gas cell. This implementation would be similar to the dust emission cycle in \textsc{skirt}, which needs some iterations due to dust self-absorption. In practice however, the relaxation time after a photo-absorption is short, and the fluorescent re-emission can be considered as being instantaneous \citep{krause}. Therefore, we implement fluorescence as a scattering process, where the incoming photon is re-emitted isotropically, with a photon energy updated to the fluorescent line energy. This allows both photo-absorption and fluorescence to be treated during \textsc{skirt}'s primary emission cycle \citep[for details, see][]{camps20}.

The effective scattering cross section for fluorescence on an atom $Z$ is just the K-shell photo-absorption cross section $\sigma_{Z,\, 1s}(E)$, multiplied by the total fluorescent yield $Y_{Z,\,K}$ of the K-shell, which is the sum of all four yields shown in Fig.~\ref{fig:yields}. The effective K-shell absorption cross section is then correspondingly updated to $(1-Y_{Z,\,K}) \, \sigma_{Z,\, 1s}(E)$, and Eq.~(\ref{eq:PA}) should be interpreted as the total extinction cross section due to photo-absorption and fluorescence combined. When fluorescent re-emission occurs in an atom, the scattered photon is assigned a fluorescent line energy sampled from the discrete probability distribution given by the set of yields $\left\{ Y_{Z,\,K_i}\right\}$. Finally, this line energy is Doppler shifted to be consistent with the thermal velocity of the atom (sampled from a Maxwell distribution), in addition to the bulk velocity of the gas, producing self-consistently broadened lines.

\subsubsection{Bound-electron scattering}
\label{sect:BES}
X-ray scattering in cold atomic gas is mainly caused by the electrons that are bound to the neutral gas atoms. Two different types of bound-electron scattering can be distinguished. Inelastic bound-Compton scattering describes the incoherent, particle-like scattering of X-ray photons on individual bound electrons, while elastic Rayleigh scattering describes the coherent, wave-like scattering on atoms as a whole. Both types of bound-electron scattering are important for modelling the X-ray spectra of obscured AGN, as these interactions make up the reprocessed X-ray continuum, and produce characteristic spectral features such as the Compton hump and the \element[][]{Fe} K$_\alpha$ Compton shoulder.

\begin{figure}
\centering
\includegraphics[width=\hsize]{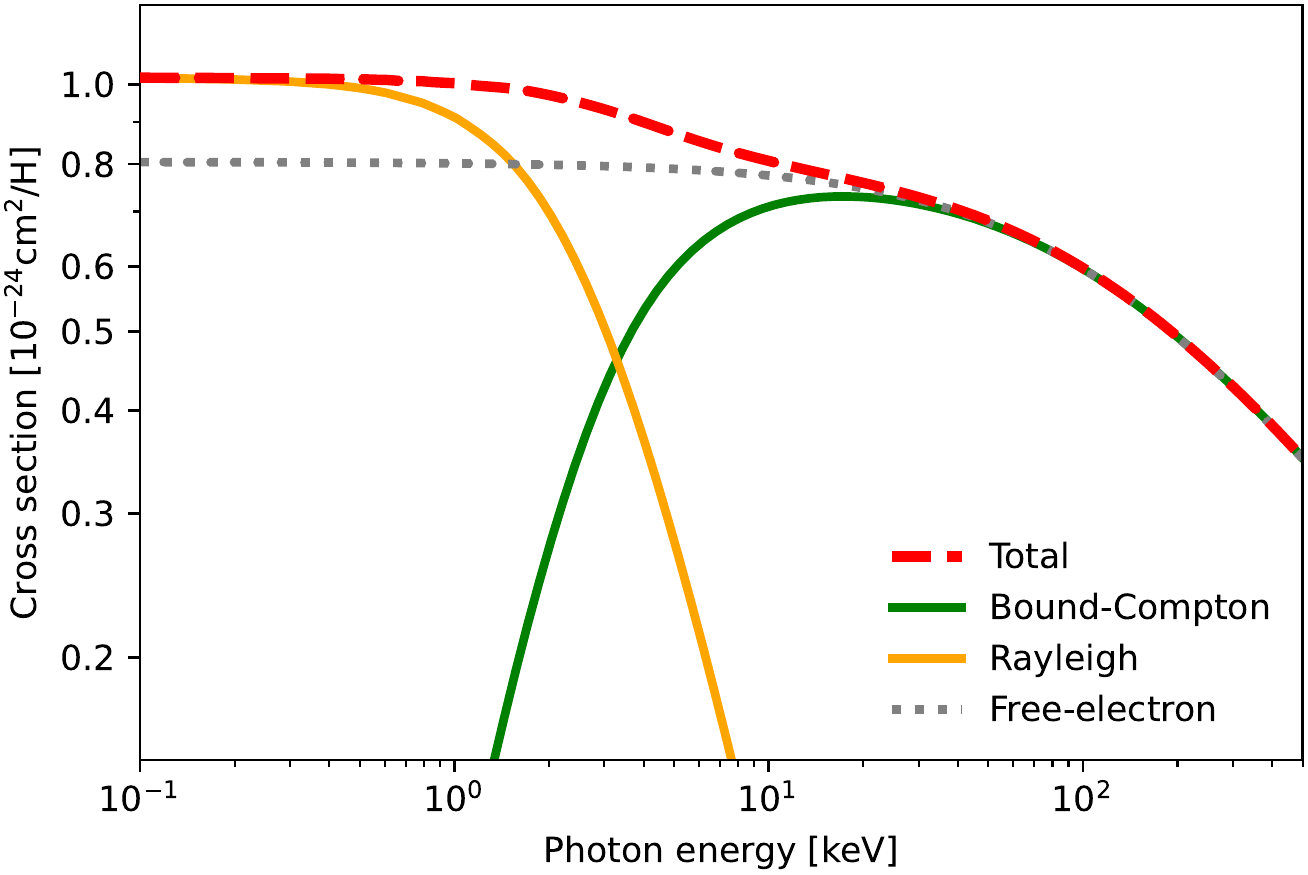}
  \caption{Bound-electron scattering cross section of a cold gas medium, assuming \citet{angr} abundances and no dust. The $1.21$ free-electron approximation is shown in grey for comparison (see text).}
     \label{fig:bes}
\end{figure}
To describe the bound-Compton scattering and Rayleigh scattering processes, we introduce the dimensionless momentum transfer parameter $q$, following the definition of \citet{Hubbell75}:
\begin{equation}
    q = \frac{E}{12.4 \, \textrm{keV}} \, \sin\left(\frac{\theta}{2}\right), \label{eq:q}
\end{equation}
which is a function of the photon energy $E$ and the scattering angle $\theta$. The differential scattering cross section for bound-Compton scattering on atom $Z$ is just the Klein-Nishina formula (Eq.~(\ref{differentialkleinnishina})), with a multiplicative correction factor $S_Z(q)$:
\begin{equation}
    \frac{d\sigma_{\text{CS},\, Z}}{d\Omega}\left(\theta, E\right)=\frac{3\, \sigma_\text{T}}{16\pi}\, \left[C^3 + C -C^2\sin^2\theta\right] \, S_Z(q),
    \label{eq:CS}
\end{equation}
where $C \equiv C(\theta, E)$ is the Compton factor, which does also depend on the scattering angle and the photon energy, see Eq.~(\ref{comptonfactor}). The correction factor $S_Z(q)$ is the incoherent scattering function, which is provided by \citet{Hubbell75} for all elements in the photo-absorbing gas (see Sect.~\ref{sect:PA}). This function represents the binding-energy correction to the free-electron description of X-ray scattering (see Sect.~\ref{sect:CS}), and is a smooth function of~$q$ that converges to $Z$ for large~$q$, and decreases to zero for small~$q$. The total scattering cross section for bound-Compton scattering on element $Z$ can be obtained as:
\begin{equation}
    \sigma_{\text{CS},\, Z}(E)= 2\pi \, \int_0^\pi d\theta \, \sin\theta \, \frac{d\sigma_{\text{CS},\, Z}}{d\Omega}\left(\theta, E\right),
\end{equation}
which needs to be evaluated numerically for each element, based on the tabulated incoherent scattering functions. The total bound-Compton scattering cross section per H-atom is then:
\begin{equation}
    \sigma_\text{CS}(E) = {\sum_{Z} \: a_Z\, \left(1-\beta_Z\right)\, \sigma_{\text{CS},\, Z}(E)},
\end{equation}
with $a_Z$ the abundance of element $Z$ relative to \ion{H}{I}, and $\beta_Z$ the correction for atoms locked up in dust grains (see Sect.~\ref{sect:PA}).
The total scattering cross section $\sigma_\text{CS}(E)$ is shown in Fig.~\ref{fig:bes}, assuming \citet{angr} abundances and no dust. Bound-Compton scattering is most important at high photon energies, where it forms the main source of X-ray extinction, see Sect.~\ref{sect:totalgasextinction}. Below $20~\text{keV}$ however, bound-Compton scattering is heavily reduced, as the scattering functions $S_Z(q)$ start suppressing the free-electron formula in Eq.~(\ref{eq:CS}). The corresponding scattering phase function is:
\begin{equation}
    \Phi_{\text{CS}}\left(\theta, E\right)= \frac{1}{\sigma_{\text{CS}}(E)} \, \sum_{Z} \: a_Z\, \left(1-\beta_Z\right)\,     \frac{d\sigma_{\text{CS},\, Z}}{d\Omega}\left(\theta, E\right),
\end{equation}
\begin{figure}
\centering
\includegraphics[width=\hsize]{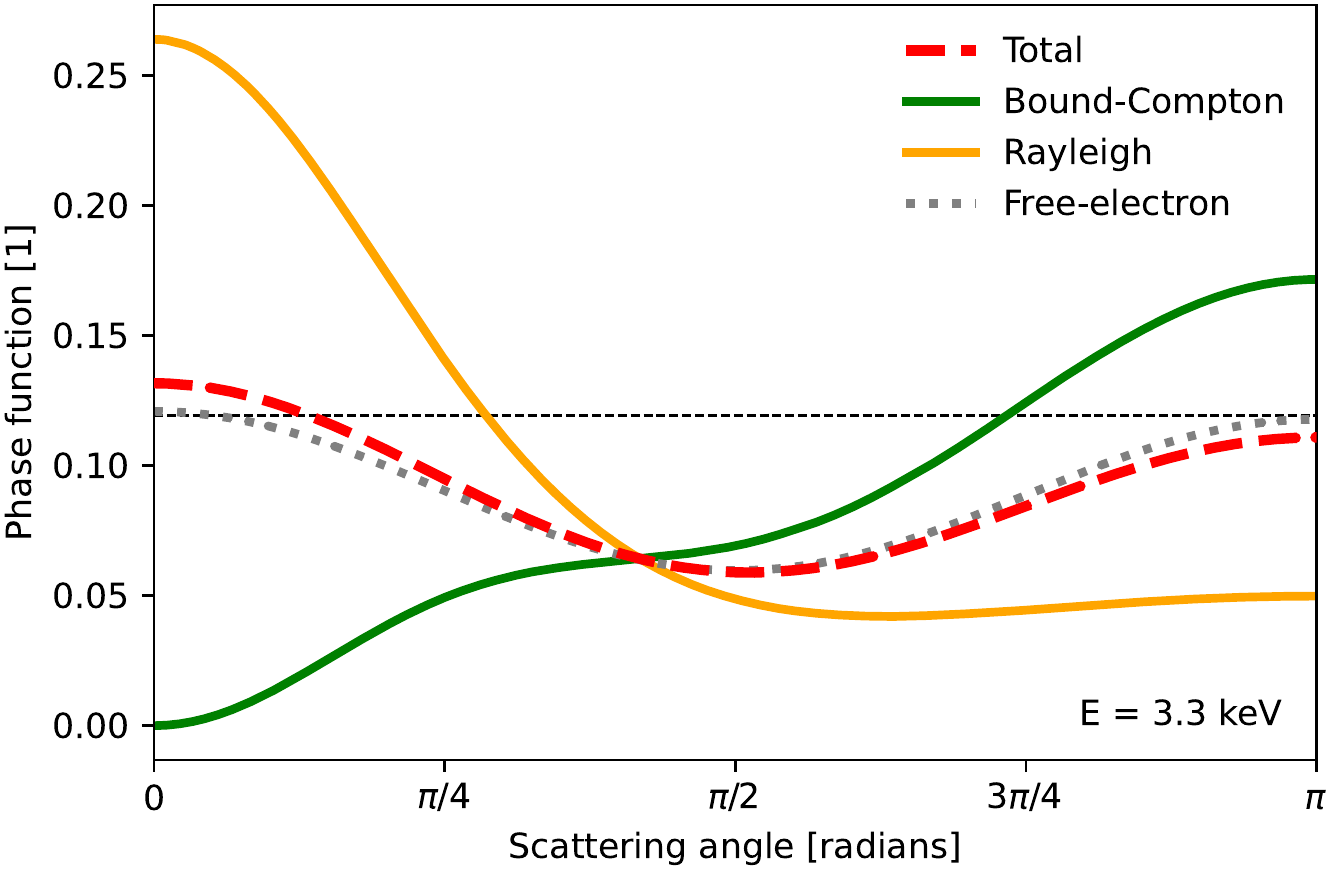}
  \caption{The normalised scattering phase functions for bound-electron scattering in cold gas, at a photon energy of $3.3~\text{keV}$. At this photon energy, bound-Compton scattering and Rayleigh scattering are equally important, see Fig.~\ref{fig:bes}. The free-electron phase function at $3.3~\text{keV}$ is shown in grey for comparison.}
     \label{fig:besphase}
\end{figure}which is illustrated in Fig.~\ref{fig:besphase} for a photon energy of $3.3~\text{keV}$. Bound-Compton scattering strongly suppresses the forward scattering direction, which was the preferred direction for free-electron scattering, see Sect.~\ref{sect:CS}. Finally, bound-Compton scattering is an inelastic process, changing the incoming photon energy as described by Eq.~(\ref{eq:ComptonEchange}).

Analogously, the differential scattering cross section for elastic Rayleigh scattering on element $Z$ is given by the Thomson formula (Eq.~(\ref{DifferentialThomson})), with a multiplicative correction factor $F^2_Z(q)$:
\begin{equation}
    \frac{d\sigma_{\text{RS},\,Z}}{d\Omega}(\theta, E)=\frac{3\, \sigma_\text{T}}{16\pi}\,\left[1 + \cos^2\theta \right] \, F^2_Z(q), \label{eq:DRS_smooth}
\end{equation}
where $F_Z(q)$ is the atomic form factor of element $Z$, which has been calculated by \citet{Hubbell75} for all elements \element[][]{H} to \element[][]{Zn}. This form factor represents the quantum mechanical scattering amplitude for X-ray scattering on a cloud of bound electrons, which converges to $Z$ at small $q$, and decreases to zero at large~$q$. The total cross section for Rayleigh scattering on $Z$ is then:
\begin{equation}
    \sigma_{\text{RS},\, Z}(E)= 2\pi \, \int_0^\pi d\theta \, \sin\theta \, \frac{d\sigma_{\text{RS},\, Z}}{d\Omega}\left(\theta, E\right),
\end{equation}
and from these tables, the total Rayleigh scattering cross section per H-atom can be calculated as:
\begin{equation}
    \sigma_{\text{RS}}(E) = {\sum_{Z} \: a_Z\, \left(1-\beta_Z\right)\, \sigma_{\text{RS},\, Z}(E)}.
    \label{eq:smoothRS}
\end{equation}
which is shown in Fig.~\ref{fig:bes}. Rayleigh scattering forms the main scattering channel at soft X-ray energies, allowing some soft X-ray photons to escape the system before photo-absorption can occur. Finally, the corresponding scattering phase function can be obtained as:
\begin{equation}
    \Phi_{\text{RS}}\left(\theta, E\right)= \frac{1}{\sigma_{\text{RS}}(E)} \, \sum_{Z} \: a_Z\, \left(1-\beta_Z\right)\,     \frac{d\sigma_{\text{RS},\, Z}}{d\Omega}\left(\theta, E\right),
\end{equation}
which is illustrated for $3.3~\text{keV}$ photons in Fig.~\ref{fig:besphase}, showing a clear bias towards forward scattering.

\begin{figure}
\centering
\includegraphics[width=\hsize]{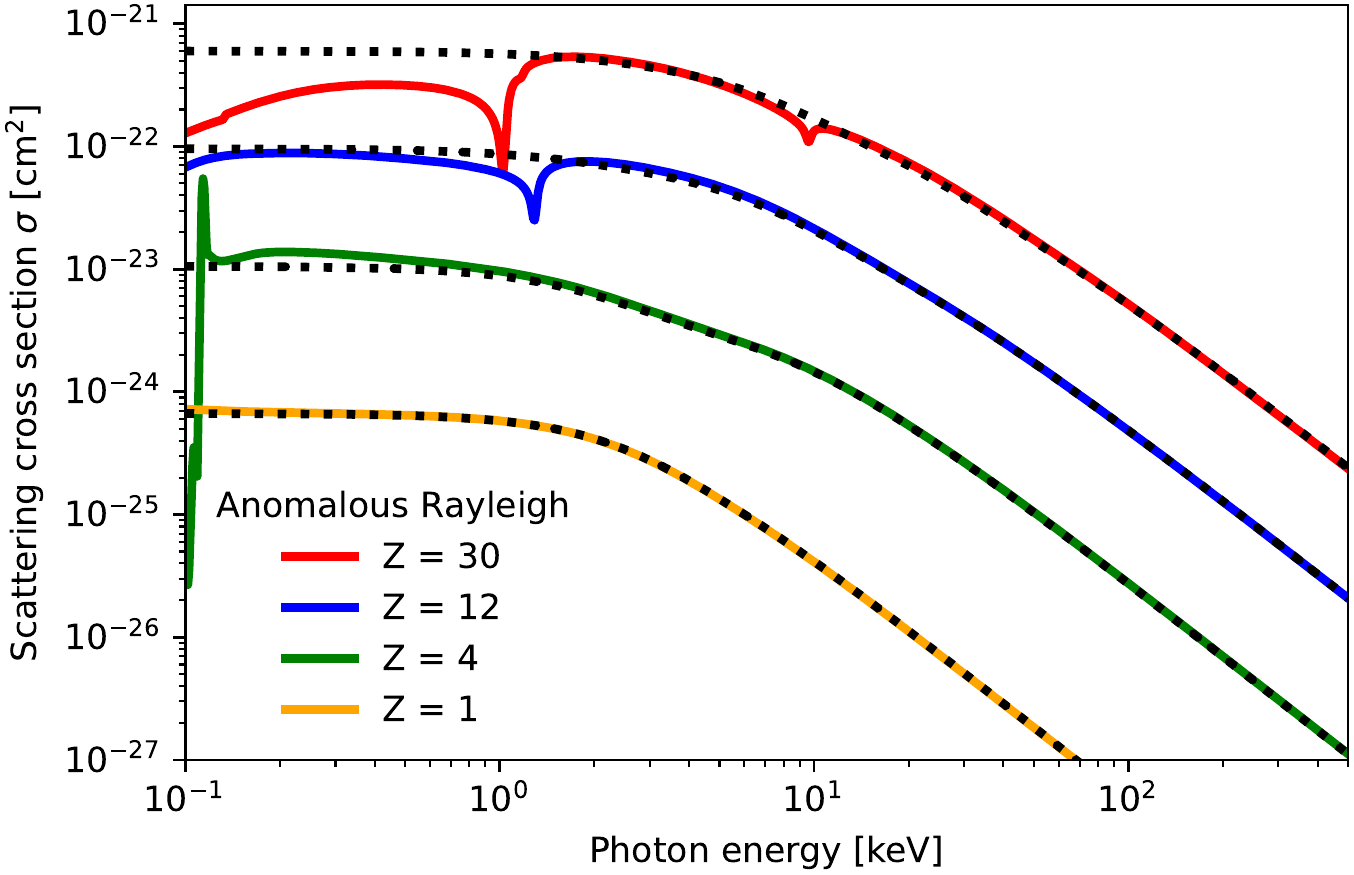}
  \caption{The anomalous Rayleigh scattering cross sections of \element[][]{H}, \element[][]{Be}, \element[][]{Mg}, and \element[][]{Zn}. The corresponding smooth Rayleigh cross sections (see Eq.~(\ref{eq:smoothRS})) are shown in black.}
     \label{fig:anomalous}
\end{figure}
A final level of refinement in treating bound-electron scattering is obtained by incorporating the anomalous scattering factors for Rayleigh scattering, which introduce spectral features close to the atomic absorption edges:
\begin{equation}
    \frac{d\sigma_{\text{RS}',\, Z}}{d\Omega}(\theta, E)=\frac{3\, \sigma_\text{T}}{16\pi}\left[1 + \cos^2\theta \right] \left[\left(F_Z(q) + F'_Z(E)\right)^2 + {F''_Z}^2(E)\right],
\end{equation}
with $F'_Z(E)$ and $F''_Z(E)$ the real and imaginary anomalous scattering factors of element Z, which are provided by \citet{cullen1997epdl97} for all 30 elements in the gas. These anomalous scattering factors correct the Rayleigh formula (Eq.~(\ref{eq:DRS_smooth})) for resonant scattering near the atomic absorption edges, and depend on the photon energy only. The anomalous Rayleigh scattering cross sections of four selected elements are shown in Fig.~\ref{fig:anomalous}, displaying distinct spectral features at the absorption edge energies. Away from the absorption edges, these cross sections converge to the original description of smooth Rayleigh scattering.

Some X-ray radiative transfer codes are approximating bound-electron scattering by free-electron scattering, which is somewhat easier to implement, see Sect.~\ref{sect:CS}. In addition, older codes might have had to consider computational constraints which did not allow for sampling the complex numerical functions that were introduced in this section. This free-electron approximation could be justified above $30~\text{keV}$, where bound-Compton scattering converges to free-electron scattering, and Rayleigh scattering is negligible, see Sect.~\ref{sect:totalgasextinction}. For \citet{angr} solar abundances, the free-electron approximation is equivalent to assuming Compton scattering on $1.21$ free electrons per \element[][]{H}-atom:
\begin{equation}
    \sum_{Z} \: a_Z\, Z = 1.21,
    \notag
\end{equation}
which is shown in Fig.~\ref{fig:bes} and Fig.~\ref{fig:besphase} for comparison. However, while this approximation matches the bound-electron scattering cross section where scattering dominates the total gas extinction (i.e.\ above $30~\text{keV}$, see Sect.~\ref{sect:totalgasextinction}), it underpredicts scattering at lower energies (see Fig.~\ref{fig:bes}), where bound-electron scattering is still important. Furthermore, the free-electron approximation ignores the important subtlety that scattering is elastic at low X-ray energies, and inelastic at high X-ray energies. In addition, this approximation does not incorporate the distinct phase function shapes that govern the X-ray scattering physics (see Fig.~\ref{fig:besphase}), and have their effect on the reprocessed X-ray emission. 

The computational efficiency of modern radiative transfer codes such as \textsc{skirt} combined with the availability of powerful computers now allows for physically more correct simulations that include bound-electron scattering. \textsc{skirt} simulations with bound-electron scattering enabled are three times slower than the corresponding free-electron simulations. However, the effect of bound-electron scattering can be observed in the resulting X-ray spectra and should not be neglected, especially in X-ray radiative transfer simulations that cover photon energies below $30~\text{keV}$. For benchmarking purposes, we provide the options to approximate bound-electron scattering by free-electron scattering, or to fully ignore scattering in cold gas. 

\subsubsection{Total gas extinction}
\label{sect:totalgasextinction}
\begin{figure}
\centering
\includegraphics[width=\hsize]{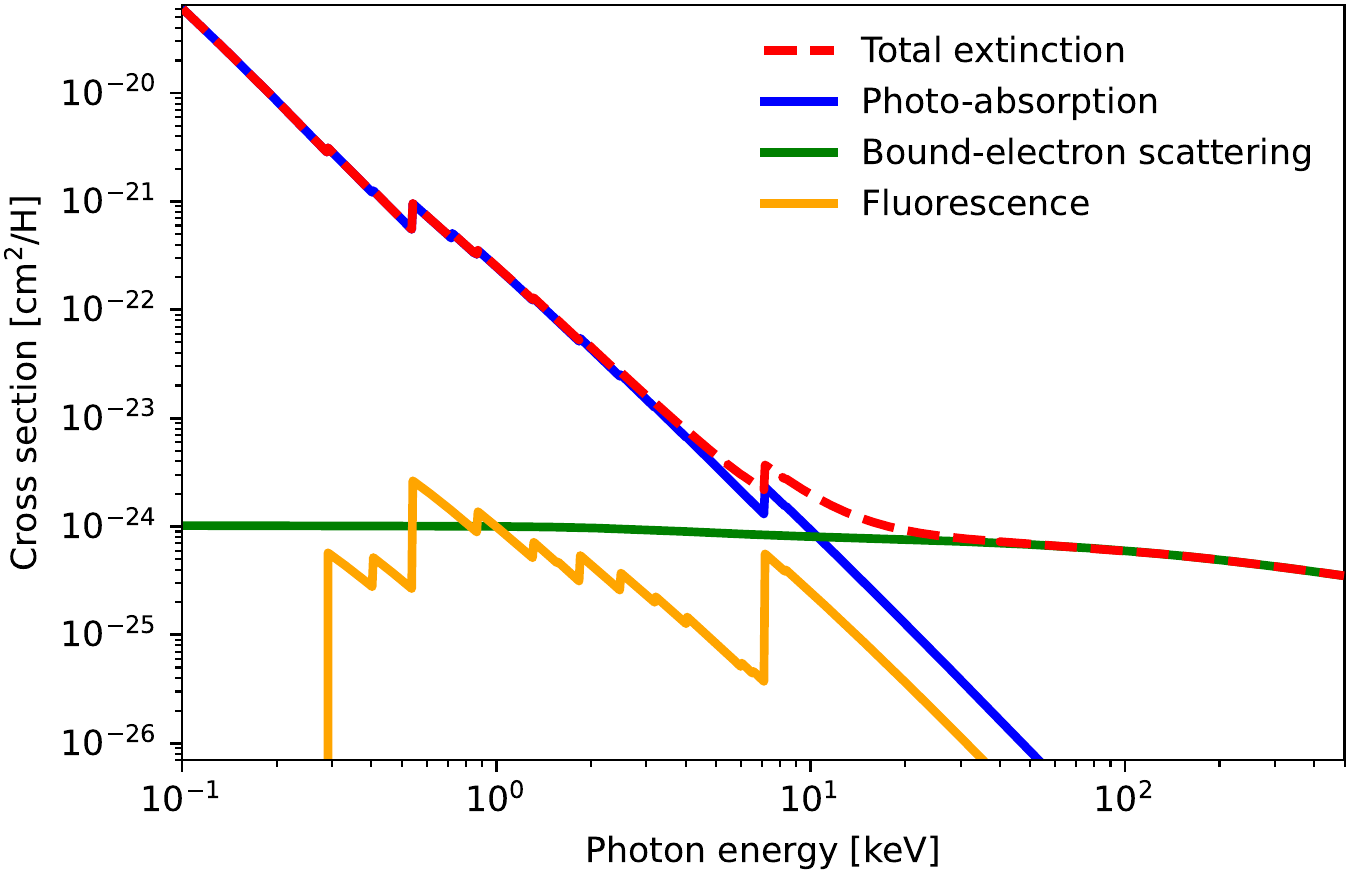}
  \caption{Total gas extinction cross section for \citet{angr} abundances and no dust. The effective photo-absorption (blue), bound-electron scattering (green), and effective fluorescent scattering (yellow) contributions are explicitly shown, to illustrate how these processes shape the X-ray spectrum in radiative transfer simulations.}
     \label{fig:totalextinction}
\end{figure}

Figure~\ref{fig:totalextinction} shows the total extinction cross section of cold gas over the $0.1$ to $500~\text{keV}$ range. Below $12~\text{keV}$, the total gas extinction is dominated by photo-absorption (Sect.~\ref{sect:PA}), which removes most soft X-ray photons from the radiation field. The effective photo-absorption cross section is shown in blue, which is Eq.~(\ref{eq:PA}) corrected for fluorescent re-emission as described in Sect.~\ref{sect:FL}. Fluorescence (yellow) has a limited contribution to the total gas extinction, but shapes the X-ray reflection spectrum by re-emitting photons at fixed line energies (Sect.~\ref{sect:BES}), producing fluorescent line features.

Above $12~\text{keV}$, bound-electron scattering (Sect.~\ref{sect:BES}) starts to dominate the total gas extinction. Fig.~\ref{fig:totalextinction} shows the bound-electron scattering cross section in green, which is the sum of bound-Compton scattering and Rayleigh scattering (see Fig.~\ref{fig:bes}). Bound-Compton scattering forms the main source of cold-gas extinction at high energy, and produces the characteristic X-ray reflection spectrum including the Compton hump and the \element[ ][]{Fe}~K$_\alpha$ Compton shoulder. Rayleigh scattering is most important at low X-ray energies, but never dominates the total gas extinction. Yet, it forms the main scattering channel for soft X-ray photons, producing the X-ray reflection component below $4~\text{keV}$ (Fig.~\ref{fig:bes}).

The X-ray reflection continua of AGN circumnuclear media are produced by Compton down-scattering on bound electrons, as described in Sect.~\ref{sect:BES}. Therefore, the observed spectrum at a given photon energy will be formed by photon interactions at higher energies, which have to be included in the radiative transfer simulations. We found that simulated X-ray spectra up to $100$ and $150~\text{keV}$ attain full convergence when considering X-ray interactions up to $200$ and $400~\text{keV}$, respectively. In addition, we found that \textsc{skirt} simulations can obtain accurate model spectra over the entire spectral range of current hard X-ray observatories such as \emph{NuSTAR} \citep{harrison13} and \emph{Swift}/BAT \citep{gehrels04, barthelmy05, krimm13}, and proposed hard X-ray missions such as \emph{HEX-P} \citep{madsen19}.

\subsection{Dust media}
\label{sect:DU}
\subsubsection{X-ray dust model}
\label{sect:xraydustmodel}
The detailed three-dimensional distribution of dusty gas plays a major role in the IR-to-UV spectral modelling of circumnuclear media, see Sect.~\ref{sect:intro}. Nevertheless, dust has mostly been neglected in X-ray radiative transfer simulations of AGN, even though it could contribute significantly to the total X-ray extinction. Furthermore, the absorption of X-ray photons could alter the dusty medium state, i.e.\ heat or destroy the dust grain population, affecting the observations at IR to UV wavelengths. Therefore, we introduce an X-ray dust model to the \textsc{skirt} code, extending the advanced treatment of dust radiative transfer into the X-ray range. This will allow for self-consistent model predictions covering both the X-ray band and the IR-to-UV wavelength range, linking X-ray reprocessing to dust modelling in AGN.

\begin{figure}
\centering
\includegraphics[width=\hsize]{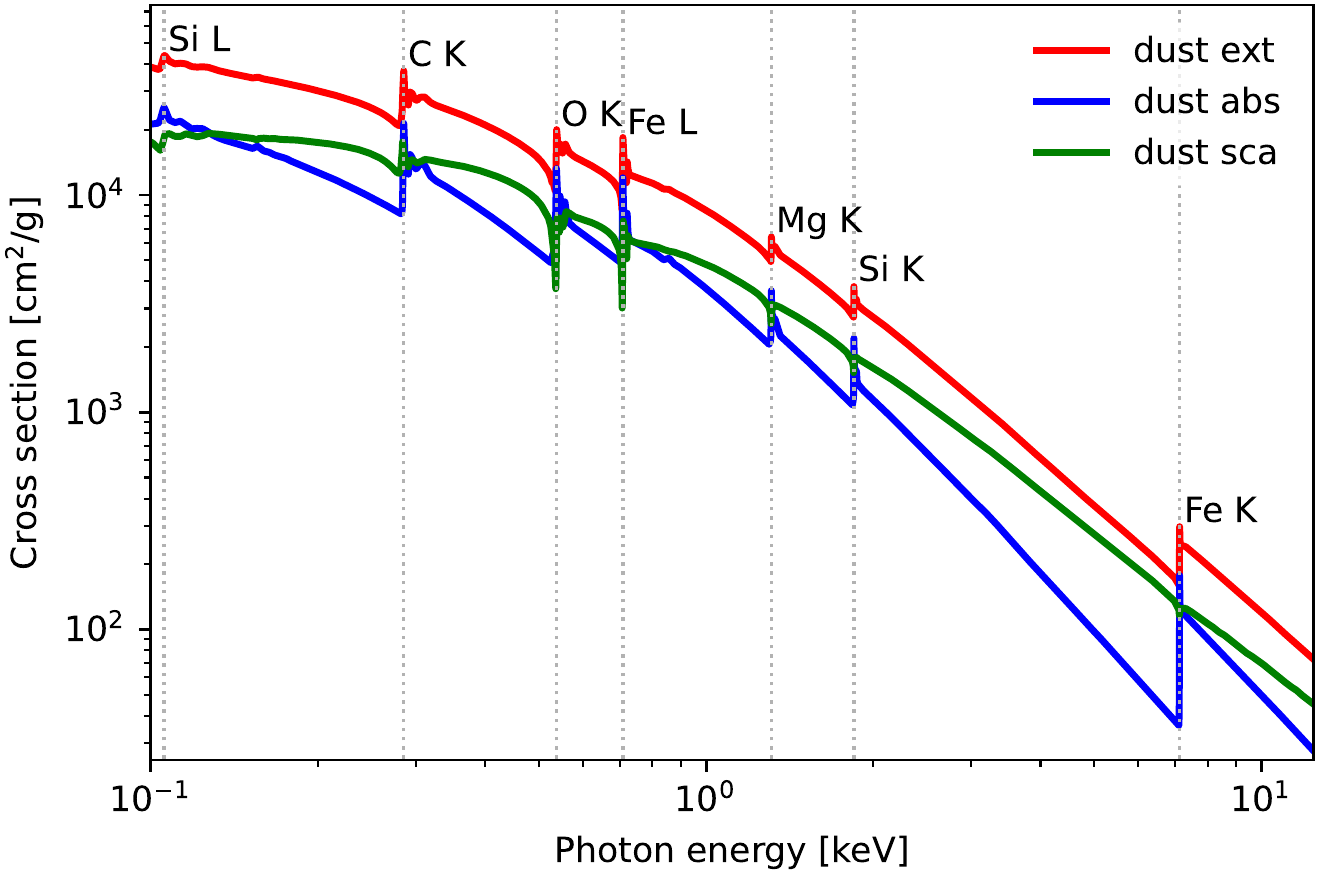}
  \caption{Dust extinction cross sections in the X-ray range, as implemented in \textsc{skirt}. The atomic absorption edges of the elements that make up the dust grains are indicated in grey.}
     \label{fig:dustcs}
\end{figure}
As a first X-ray dust model for \textsc{skirt}, we implement a dust mix of carbonaceous grains and amorphous silicate grains with a \citet{WD01} grain size distribution. This dust model has already been implemented in \textsc{skirt} for IR to UV wavelengths \citep[][]{li01}, allowing the dust mix to be used over the entire $10^{-4}$ to $10^{4}~\mu\text{m}$ range. The X-ray extinction cross sections for this model were computed\footnote{\url{https://www.astro.princeton.edu/~draine}} by \citet{draine03} for photon energies between $0.1~\text{keV}$ and $12.4~\text{keV}$, and represent the absorption and scattering probabilities of individual dust grains that are representative for the entire dust mix. Absorption by dust grains is similar to photo-absorption by the neutral gas atoms that constitute the dust mix composition (Sect.~\ref{sect:PA}), but accounts for shielding at soft X-ray energies and interference in the solid grains. Scattering on dust grains is an elastic process, which was modelled using Mie theory or anomalous diffraction theory depending on the photon energy \citep[see][for more details]{draine03}. These cross sections are shown in Fig.~\ref{fig:dustcs}, and contain detailed spectral features near the atomic absorption edges, which form an interesting diagnostic on the dust composition, crystallinity, shape, and grain size, and have been observed both in astrophysical systems and lab measurements \citep[][]{corrales16, costantini19, zeegers19, rogantini20, psaradaki20}. More advanced dust models with configurable size distributions can easily be introduced to \textsc{skirt}, owing to the modular design of the code.

Finally, a dust mass density $\rho_\text{d}(\vec{r})$ characterises the spatial distribution of the dusty medium. Often however, it is more meaningful to connect the distribution of dust to the distribution of the gas in which the dust resides, by declaring a dust mass per \element[][]{H}-atom $m_\text{d}$, so that:
\begin{equation}
    \rho_\text{d}(\vec{r}) = m_\text{d} \; n_\text{H}(\Vec{r}).
\end{equation}
For a realistic dust-to-metal mass ratio of $0.3$ \citep{HSPP1} and \citet{angr} solar abundances, this dust mass $m_\text{d}$ would be $1.38 \times 10^{-26}~\text{g}$ per \element[][]{H}-atom.

In turn, the atomic gas abundances should be updated to account for the depletion of elements that are locked up in dust grains, see Eq.~(\ref{eq:PA}). The \citet{WD01} dust mixture implemented in \textsc{skirt} consists of $84\%$ graphite and $16\%$ olivine (i.e.\ the relative number of structural units \element[][]{C} and \element[][]{MgFeSiO_4}, corresponding to dust mass fractions of $27\%$ and $73\%$, respectively). Defining the graphite abundance $a_\text{gr}$ as the number of structural units per \element[][]{H}-atom, one has:
\begin{equation}
    a_\text{gr} = \frac{0.84 \, m_\text{d}}{0.84 \, \mu_\text{gr} + 0.16 \, \mu_\text{ol}},
\end{equation}
with $\mu_\text{gr} = 12~\text{amu}$ the mass per structural unit graphite, and $\mu_\text{ol} = 172~\text{amu}$ the mass per structural unit olivine. The carbon depletion factor is then calculated as $\beta_\text{C} = a_\text{gr}/a_\text{C}$. Equivalently, the gas abundances of \element[][]{Mg}, \element[][]{Fe}, and \element[][]{Si} are reduced by the abundance of olivine:
\begin{equation}
    a_\text{ol} = \frac{0.16 \, m_\text{d}}{0.84 \, \mu_\text{gr} + 0.16 \, \mu_\text{ol}},
\end{equation}
while the oxygen depletion is calculated as $\beta_\text{O} = 4 \, a_\text{ol}/a_\text{O}$. For solar gas abundances, $m_\text{d}$ should be limited to $1.38 \times 10^{-26}~\text{g}$ per \element[][]{H}-atom, when all available \element[][]{Si} will be locked up in dust grains. Phenomenological dust masses exceeding this value could then be explained by the existence of elongated or porous dust grains, which provide more extinction than an equivalent dust mass in compact spherical grains, as implemented in our dust model \citep{reviewDraine03}.

The scattering phase function for the X-ray dust model in \textsc{skirt} is approximated by the \citet{HenyeyGreenstein} (HG) function:
\begin{equation}
\label{eq:HG}
\Phi_\mathrm{HG}(\theta, E) = \frac{1}{4 \pi} \frac{1-g^2(E)}{\left(1+g^2(E)-2g(E)\cos\theta\right)^{3/2}},
\end{equation}
which has also been used to model dust scattering in the IR-to-UV wavelength range \citep{DraineIRUV}. The $\Phi_\text{HG}$ function shape is set by an energy-dependent asymmetry parameter $g(E)$, which can be interpreted as the expected value for the cosine of the scattering angle: $-1 \le g = \left<\cos\theta\right> \le 1$. In the X-ray range, we adopt a value for $g$ in function of the photon energy $E$:
\begin{equation}
    \label{eq:g}
    g(E) = 1 - \frac{1.23 \times 10^{-3}~\mathrm{keV}}{E},
\end{equation}
which was obtained by fitting a HG function to the dust scattering phase function $\Phi_\text{D03}$ proposed by \citet{draine03}, for each photon energy. The resulting functional form $g(E)$ then matches the tabulated $g$ values adopted by \textsc{skirt} in the mm to far-UV wavelength range at $0.1 \, \text{keV}$, and causes the $\Phi_\text{HG}$ and $\Phi_\text{D03}$ phase functions to coincide at $\theta = 0$ for all photon energies.

X-ray scattering by dust can be observed directly in diffuse scattering halos around galactic X-ray point sources, which form a promising diagnostic on the dust composition and chemistry \citep[][]{DraineTan03, corrales15, corrales17, sguera20, lamer21, landstorfer22}. These dust scattering halos have all been modelled with simple one-dimensional dust screens \citep[see e.g.][]{costantini22}. With \textsc{skirt}, more complex dust geometries can now be explored, and tested against X-ray observations.

\subsubsection{Extreme forward scattering by dust}
\label{sect:extremeforwardscattering}
As noted in Sect.~\ref{sect:xraydustmodel}, the current X-ray dust model in \textsc{skirt} describes scattering on dust grains using the HG phase function $\Phi_\mathrm{HG}$ (Eq.~(\ref{eq:HG})), with asymmetry parameter values $g(E)$ given by Eq.~(\ref{eq:g}). From the latter equation, it is clear that $g$ closely approaches unity in the X-ray regime, causing the corresponding HG function to model extreme forward scattering, with $\left<\cos\theta\right> = 0.9988$ at $1~\mathrm{keV}$, and $\left<\cos\theta\right> = 0.9999$ at $10~\mathrm{keV}$. Evaluating the HG function or sampling scattering angles from this function becomes numerically unstable for large $g > 1-10^{-6}$. However, these values of $g$ are only reached for energies $E>1~\mathrm{MeV}$, well above the considered simulation domain. Nevertheless, the \textsc{skirt} implementation clips any $g$ value to the stated numerical limit, to avoid numerical instabilities with future dust mixes that could reach this limit at a lower photon energy.

More problematically, this extreme forward scattering causes issues with the scattering peel-off technique, which forms a core element of \textsc{skirt}'s radiative transfer method, as described in \citet{baes11}. Specifically, the bias factor $w_\mathrm{obs}$ for peel-off photon packets sent to an observer will become extremely large when the scattering angle towards that observer is small, as the sharply-peaked HG function yields large values for small angles, see e.g.\ Eq.~(30) of \citet{MCRT2}. As a result, some peel-off photon packets will carry an extreme, artificially-boosted amount of energy, which causes unacceptable noise levels in the fluxes measured by \textsc{skirt} \enquote{instruments} (which are hereafter defined as apertures recording photon packets leaving the system in a preset observer direction).

The bias factor for HG scattering relative to the bias factor for isotropic scattering is:
\begin{equation}
    b(\theta, E) = \frac{w_\mathrm{obs,\,HG}(\theta, E)}{w_\mathrm{obs,\,iso}}
    = \frac{\Phi_\mathrm{HG}(\theta, E)} {\Phi_\mathrm{iso}}
    = 4\pi \; \Phi_\mathrm{HG}(\theta, E),
    \label{eq:b}
\end{equation}
which equals a renormalised form of the HG function. For a given photon energy, the largest possible value of $b$ corresponds to peel-off photon packets emitted to an observer in the original photon propagation direction, so that $\theta_\mathrm{obs}=0$. Experiments show that $b$ should be kept under $\approx 10^3$, to achieve acceptable noise levels in \textsc{skirt} simulations with a manageable number of photon packets. However, for photon energies of $12.4~\mathrm{keV}$, we find $b(\theta=0)\approx 2.0 \times 10^{8}$, which is many orders of magnitude higher. Even for $0.1~\mathrm{keV}$ photons, we find $b(\theta=0)\approx 1.3 \times 10^{4}$. Thus, this peel-off problem exists for dust scattering over the entire X-ray range under consideration.

\begin{figure}
\centering
\includegraphics[width=\hsize]{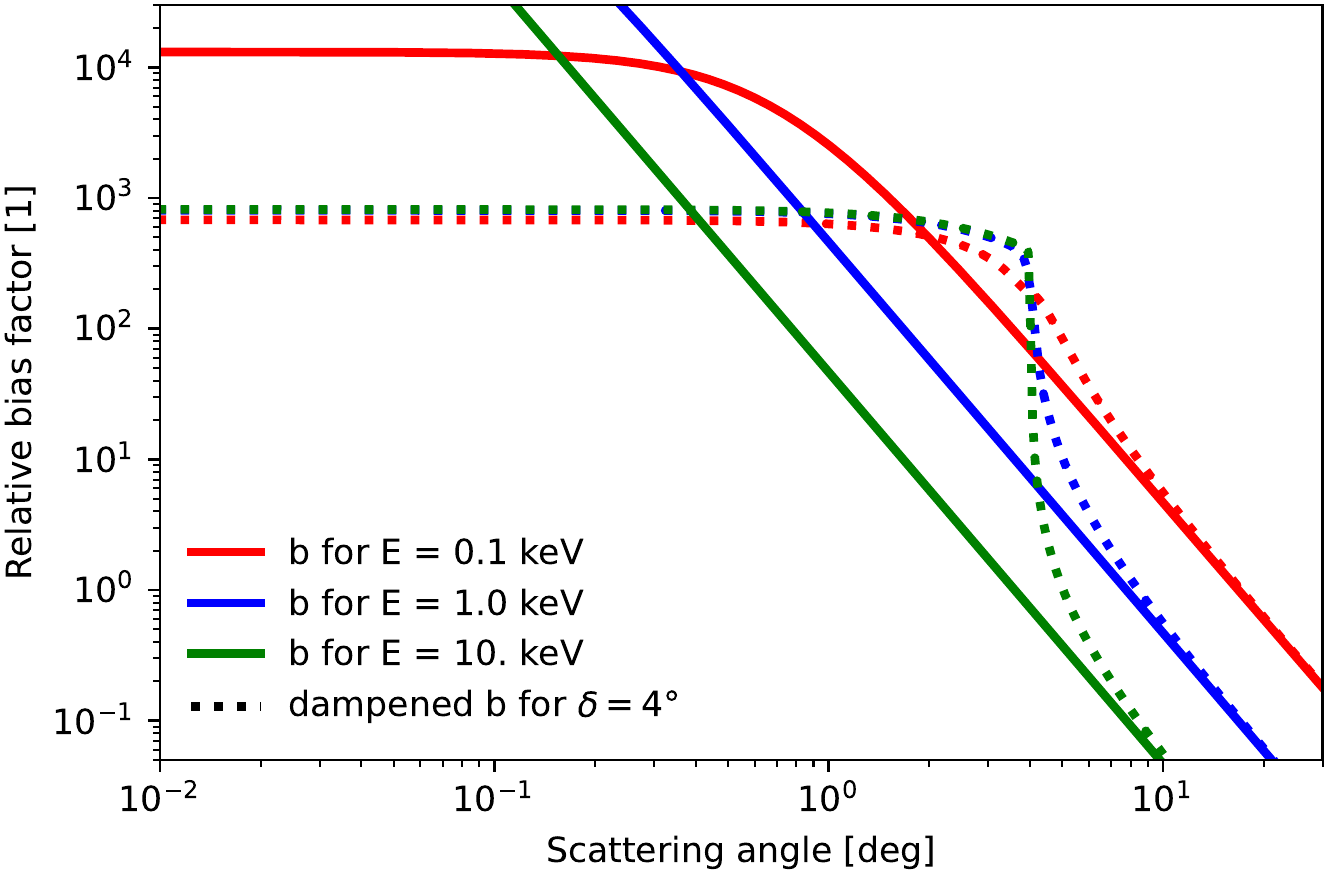}
  \caption{The dampened relative bias factors $\left<b\right>$ for dealing with extreme forward scattering by dust. For $\delta = 4^\circ$, we maintain $\left<b\right>\lesssim 820$.}
     \label{fig:dampened}
\end{figure}To address this situation, we average the phase function part of the bias factor $w_\mathrm{obs}$ over a small patch of sky around the observer direction $\vec{k}_\mathrm{obs}$, as opposed to evaluating the phase function $\Phi_\mathrm{HG}$ at $\theta_\mathrm{obs}$ exactly. More precisely, we replace $b$ by its average value $\left<b\right>$ over a small solid angle $\Omega_\mathrm{obs}$ around $\vec{k}_\mathrm{obs}$\footnote{We omitted the expressions for the special cases where the opening angle straddles $\theta=0$; these are readily derived by analogy.}:
\begin{equation}
    \left<b\right>(\theta_\mathrm{obs}, E, \delta) =\frac{\displaystyle\int_{\cos(\theta_\mathrm{obs}+\delta)}^{\cos(\theta_\mathrm{obs}-\delta)} 4\pi \; \Phi_\mathrm{HG}(\cos\theta, E) \;\mathrm{d}\cos\theta} {\cos(\theta_\mathrm{obs}-\delta)-\cos(\theta_\mathrm{obs}+\delta)},
    \label{eq:average_b}
\end{equation}
where $\theta_\mathrm{obs}$ is the angle between the original photon direction and the observing instrument, and $\delta$ can be interpreted as the half-opening angle of the instrument in the polar direction. Note how averaging in the azimuthal direction has no net effect, as the HG phase function does not depend on $\varphi$.

In the polar direction, we adopt a parameter value of $\delta = 4^\circ$. Indeed, this value for $\delta$ dampens the relative bias factor to $\left<b\right>\lesssim 820$ for all $E$, comfortably below the empirical limit of $\approx 10^3$ that was stated above. Smaller $\delta$ values would still yield bias factors at or above the empirical limit, causing unacceptable noise levels in the recorded fluxes. On the other hand, larger $\delta$ values would unnecessarily increase the artificial spatial blur in the results (see next paragraph). For the selected value of $\delta = 4^\circ$, the averaged relative bias factor $\left<b\right>$ is shown for different photon energies in Fig.~\ref{fig:dampened}, which is strongly dampened for small scattering angles, and converges to its original value for large scattering angles. The expression for averaging the bias factor (Eq.~(\ref{eq:average_b})) can be evaluated analytically at each scattering peel-off. Therefore, this implementation of extreme forward scattering is equally efficient as the original treatment without averaging.

While decreasing the noise levels as desired, this averaging process also introduces a certain amount of unwanted spatial blur to the recorded fluxes. Indeed, photons moving in a direction close to the line of sight of some instrument will be smeared out over a small solid angle, when scattering on a dust grain towards that instrument. However, this blurring is not really an issue when comparing simulation results to spatially-resolved observations of dust scattering, as it will only influence the scattered flux, mainly in the most central regions\footnote{Note how modifying the peel-off angle $\theta_\mathrm{obs}$ with $\delta$ does not equally change the corresponding halo size $\theta_\mathrm{halo}$ on the sky, as $\theta_\mathrm{halo} \approx x\, \theta_\mathrm{obs}$, with $x$ the relative distance of a dust screen between source and observer.} of extended sources, that are probably dominated by the direct source emission anyway. Furthermore, these central regions do also feature some spatial blur in actual observations, as described by the point spread function. In any case, \textsc{skirt} will properly recover the extended halo emission away from the source centre, which could be used to model halo profiles in dust studies. For X-ray sources that cannot be spatially resolved (e.g.\ AGN), the effect of spatial averaging is negligible, as the observed fluxes already entail a spatial integration over the source. Finally, we recall that the averaging procedure is only invoked at the last scattering event before a photon detection (i.e.\ at the scattering peel-off), and that all other photons performing a random walk inside the dusty model geometry are treated exactly, i.e.\ based on the unmodified, sharply-peaked HG phase function.

\section{Benchmark tests}
\label{sect:benchmark}
To verify our X-ray implementation, we compare the \textsc{skirt} simulation output\footnote{Using the publicly-available \textsc{skirt} code, git commit \texttt{6b8531e}.} with some well-established spectral models that deal with X-ray obscuration and reflection in the context of AGN. In Sect.~\ref{sect:xspecmodels}, we reproduce the setup of some one-dimensional \textsc{xspec} models for free-electron, cold-gas, and dust extinction. In Sect.~\ref{sect:torusmodels}, we benchmark our code against some popular smooth torus models that are frequently used to model observational X-ray data. In all simulations, the primary X-ray source has a power law shape with a photon index of $\Gamma=1.8$, an exponential cut-off at $E_\textrm{cut}=200~\text{keV}$, and an integrated $2-10~\text{keV}$ luminosity of $L_{2-10}=2.8 \times 10^{42}~\text{erg}~\text{s}^{-1}$, which is representative for the nearby AGN in Circinus \citep[see e.g.][]{wada16}.

\subsection{One-dimensional \textsc{xspec} models}
\label{sect:xspecmodels}

\subsubsection{\textsc{phabs} and \textsc{cabs}}
\label{sect:phabscabs}
\textsc{phabs} and \textsc{cabs} are two one-dimensional \textsc{xspec} models, dealing with cold-gas and free-electron extinction \citep[][]{arnaud96}. \textsc{phabs} models photo-absorption (Sect.~\ref{sect:PA}) by a 1D slab of neutral gas atoms, evaluating the 1D transmission function:
\begin{equation}
    T(E) = e^{-N_\text{H} \, \sigma_\text{PA}(E)},
\end{equation}
with $N_\text{H}$ the hydrogen column density of the slab, and $\sigma_\text{PA}(E)$ the photo-absorption cross section per \element{H}-atom (Eq.~(\ref{eq:PA})). \textsc{skirt} implements the exact same photo-absorption cross sections as those set by the \textsc{xspec} command \textsc{xsect vern}, allowing for a direct comparison. In addition, we can choose the same gas abundances as those set by the \textsc{xspec} command \textsc{abund angr}, i.e.\ \citet[][]{angr} abundances for all 18 elements implemented in \textsc{phabs}. The \textsc{phabs} model deals with photo-absorption only, ignoring fluorescence (Sect.~\ref{sect:FL}) and bound-electron scattering (Sect.~\ref{sect:BES}) in the gas.

\textsc{cabs} models the extinction of a 1D slab of free electrons (Sect.~\ref{sect:CS}), ignoring the actual Compton-scattered photons. This simplified treatment is valid in optically-thin media, and comes down to evaluating the 1D transmission function:
\begin{equation}
    T(E) = e^{-N_\text{e} \, \sigma_\text{KN}(E)},
\end{equation}
with $N_\text{e}$ the electron column density of the slab, and $\sigma_\text{KN}(E)$ the Compton scattering cross section (Eq.~(\ref{totalcompton})). The electron column density of the \textsc{cabs} model is often related to the hydrogen column density of an other \textsc{phabs} model as $N_\text{e} = 1.21 \times N_\text{H}$, forming a crude approximation to bound-electron scattering in a cold gas medium, see Sect.~\ref{sect:BES}. In this subsection, we will combine the \textsc{phabs} and \textsc{cabs} models to represent this approximation.

We mimic the \textsc{xspec} model setup in \textsc{skirt} with a three-dimensional slab of cold gas and free electrons, so that an X-ray point source is observed through the slab. We disable bound-electron scattering, which is a configurable option of the cold-gas mix in \textsc{skirt} (see Sect.~\ref{sect:BES}). We run \textsc{skirt} simulations for different values of $N_\text{H}$, each time linking the electron column density to the hydrogen column density as $N_\text{e} = 1.21 \times N_\text{H}$. As the \textsc{phabs} and \textsc{cabs} models ignore both fluorescent line emission and scattered photons, we focus on the direct observed flux only, i.e.\ transmitted photons which did not interact within the slab. It is possible to retrieve this individual flux component using the \enquote{smart} photon instruments in \textsc{skirt}, which can access in-simulation information such as the interaction history of detected photons, which is not available in actual observations.

\begin{figure}
\centering
\includegraphics[width=\hsize]{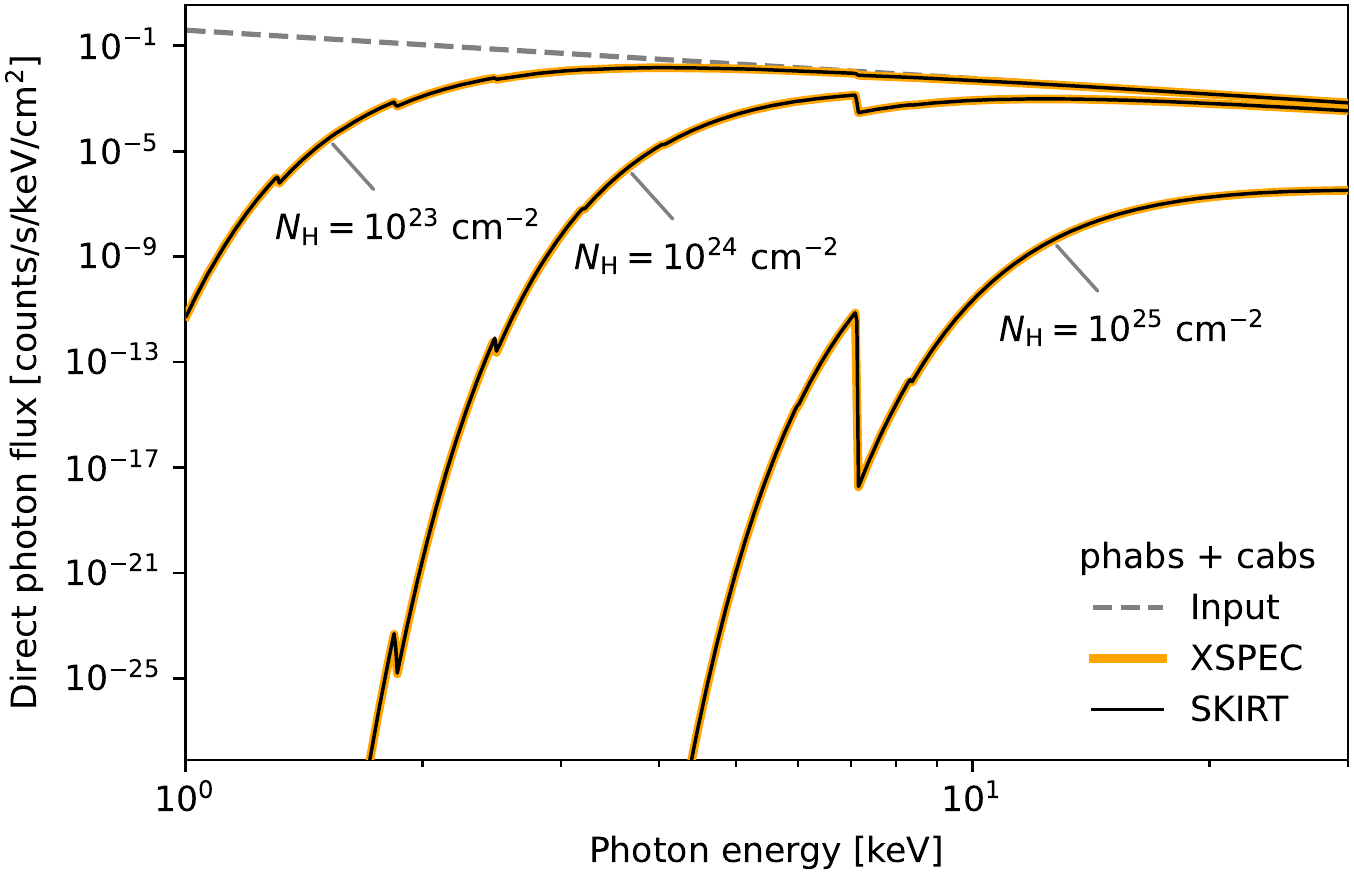}
  \caption{First \textsc{skirt} benchmark (Sect.~\ref{sect:phabscabs}), comparing three \textsc{skirt} simulations with the \textsc{phabs} and \textsc{cabs} models for cold-gas and free-electron extinction. The hydrogen column density $N_\text{H}$ is indicated on the figure, while the electron column density is $N_\text{e} = 1.21 \times N_\text{H}$ (see text).}
     \label{fig:phabscabs}
\end{figure}
The \textsc{skirt} simulation output is shown in Fig.~\ref{fig:phabscabs}, in excellent agreement with the corresponding \textsc{xspec} results. And even though the \textsc{phabs} and \textsc{cabs} models do not incorporate any non-trivial radiative transfer effects, they assure a correct implementation of the basic radiative processes in free-electron and cold-gas media. Equivalently, we ran \textsc{skirt} simulations with cold gas only, this time enabling the free-electron approximation for bound-electron scattering in the gas (see Sect.~\ref{sect:BES}). In these simulations, we obtained the exact same results as the ones shown in Fig.~\ref{fig:phabscabs}, in agreement with the underlying assumption of scattering on 1.21 free electrons per \element[][]{H}-atom.

\subsubsection{\textsc{pexmon}}
\label{sect:pexmon}
 \textsc{pexmon} \citep{nandra07} is the standard \textsc{xspec} model for dealing with reflection by neutral material, which combines the \textsc{pexrav} \citep{magdziarz95} continuum reflection model with self-consistently generated Fe K$_\alpha$, \element[][]{Fe} K$_\beta$, and \element[][]{Ni} K$_\alpha$ fluorescent lines, plus a \element[][]{Fe} K$_\alpha$ Compton shoulder. For a \textsc{pexmon} model parameter $\textsc{rel\_refl} = -1$, this model assumes a semi-infinite slab of optically-thick atomic gas, illuminated by an X-ray point source. The reflected X-ray flux is then observed from the same side of the slab, for different orientations with respect to the reflecting plane. Inside the gas, all \element[][]{H} and \element[][]{He} atoms are assumed to be fully ionised, and Compton scattering (Sect.~\ref{sect:CS}) on these free electrons produces the observed reflection continuum. All other elements are assumed to be neutral, causing photo-absorption similar to the \textsc{phabs} model (Sect.~\ref{sect:phabscabs}). In addition, three fluorescent lines are incorporated into the model (Fe K$_\alpha$, \element[][]{Fe} K$_\beta$, and \element[][]{Ni} K$\alpha$), based on calculations by \citet[][]{george91}. Bound-electron scattering is ignored for all elements in the gas.

We mimic the \textsc{pexmon} model setup in \textsc{skirt} using a three-dimensional slab of cold gas and free electrons, with an X-ray point source close to the slab, and an observer at the same side. We set \citet[][]{angr} abundances for all elements in the photo-absorbing gas, except for \element[][]{H} and \element[][]{He} which are set to zero. Furthermore, we disable any kind of scattering by the gas. We link the density of the free-electron medium to the cold gas as $n_\text{e} = 1.20 \times n_\text{H}$, which corresponds to a free-electron population formed by ionising all \element[][]{H} and \element[][]{He} atoms in a \citet[][]{angr} gas mix. We set the hydrogen column density\footnote{Note how the distribution of photo-absorbing gas atoms is defined relative to a representative neutral hydrogen density, even when the actual neutral hydrogen abundance of the gas mix is zero, see Sect.~\ref{sect:PA}.} through the slab to an arbitrary high value of $N_\text{H} = 10^{27}~\text{cm}^{-2}$, and the corresponding electron column density to $N_\text{e} = 1.20 \times 10^{27}~\text{cm}^{-2}$. These high column densities result in an optical depth through the slab of $\tau>500$ over the entire simulation domain, which makes the slab virtually impenetrable, modelling an optically-thick semi-infinite slab of cold material.

The total observed flux for this model consists of a direct source component, plus a contribution of scattered photons, i.e.\ photons which interacted at least once with the transfer medium. In this subsection, we focus on the scattered component only, as only scattered photons can reveal the spectral effect induced by radiative transfer processes. Spectra of scattered photons can be retrieved as separate components with both \textsc{pexmon} and \textsc{skirt}, which can be compared to benchmark the \textsc{skirt} implementation. Note how fluorescent line photons are considered to be part of the scattered flux component, adding lines to the X-ray reflection spectrum.

\begin{figure}
\centering
\includegraphics[width=\hsize]{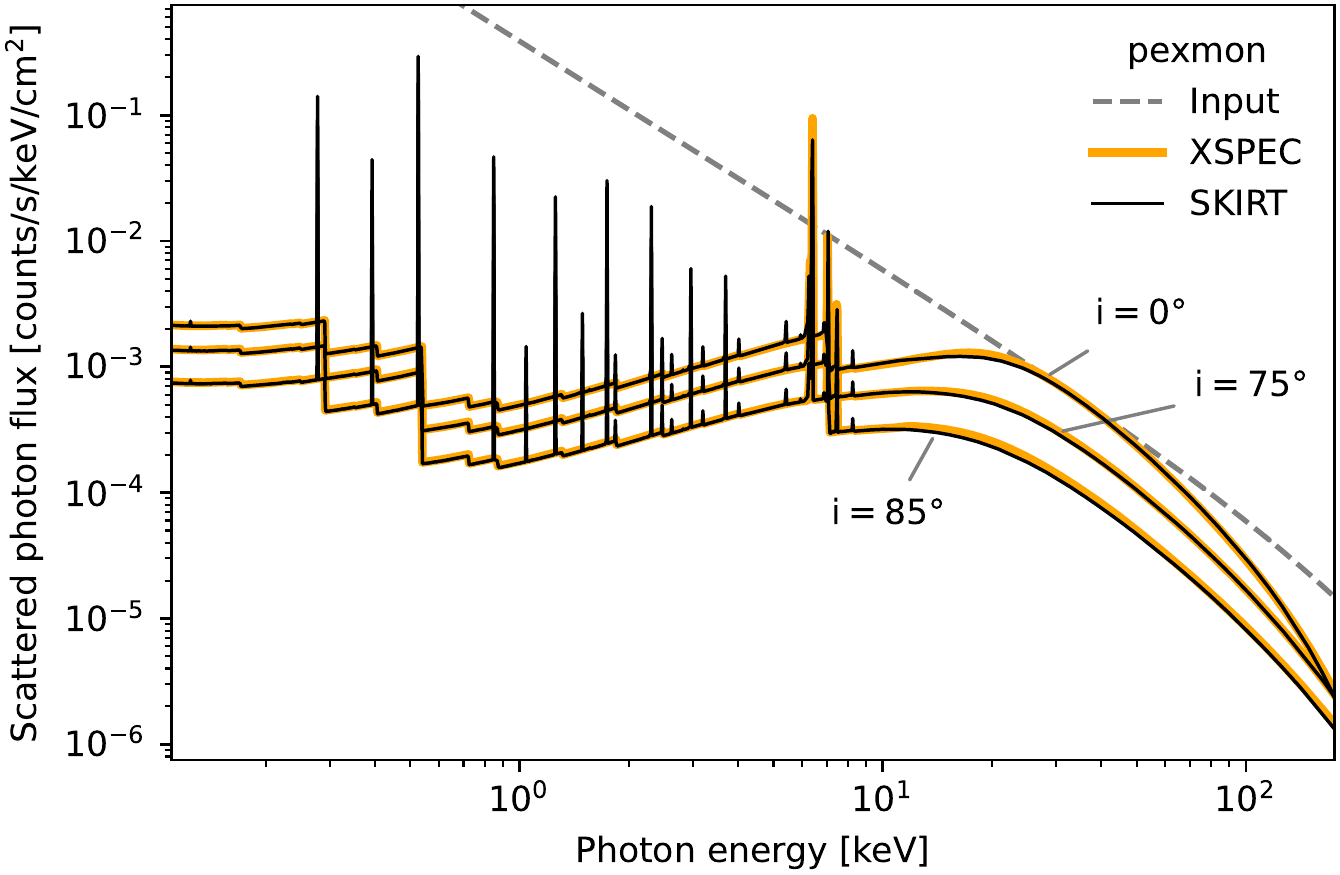}
  \caption{Second \textsc{skirt} benchmark (Sect.~\ref{sect:pexmon}), comparing three \textsc{skirt} simulations with the \textsc{pexmon} model for reflection by neutral material. The inclination relative to the reflecting plane is indicated on the figure. The $10\%$ discrepancy in the $12$ to $60~\text{keV}$ range was expected (see text).}
     \label{fig:pexmon}
\end{figure}
Fig.~\ref{fig:pexmon} shows the \textsc{skirt} simulation results for the \textsc{pexmon} model setup, for three different observing angles $i$ relative to the reflection-plane normal. The \textsc{skirt} output spectra show many more fluorescent lines, which is expected as \textsc{skirt} implements four line transitions for each element in the photo-absorbing gas (Sect.~\ref{sect:FL}). The reflected X-ray continuum is found to be in excellent agreement with the \textsc{pexmon} model over the entire simulation domain, except for the $12$ to $60~\text{keV}$ range, where \textsc{pexmon} overestimates the scattered flux by up to $10\%$. However, this behaviour of the underlying \textsc{pexrav} reflection model was identified by \citet{magdziarz95}, and is therefore expected.

\subsubsection{\textsc{ismdust}}
\label{sect:ismdust}
\begin{figure}
\centering
\includegraphics[width=\hsize]{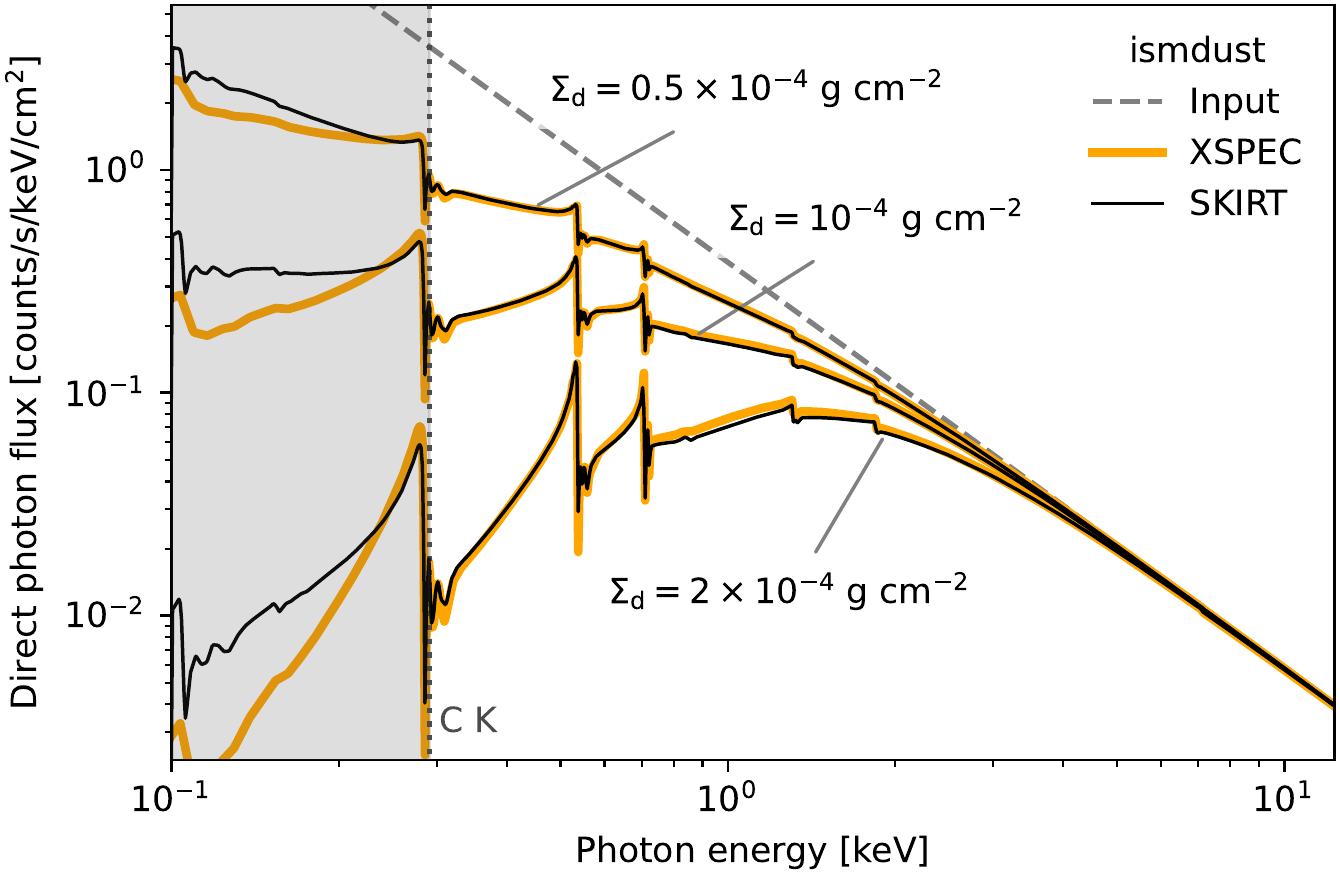}
  \caption{Third \textsc{skirt} benchmark (Sect.~\ref{sect:ismdust}), comparing \textsc{skirt} with the \textsc{ismdust} model for dust extinction. The dust mass column density $\Sigma_\text{d}$ is indicated on the figure. Both models are in good agreement above the \element[][]{C}~K-edge at $0.280~\text{keV}$.}
     \label{fig:ismdust}
\end{figure} \textsc{ismdust} is a one-dimensional \textsc{xspec} model for dust extinction, modelling a 1D slab of graphite and olivine dust grains \citep[][]{corrales16}. This dust model is based on the same optical dust properties as those incorporated in \textsc{skirt}, but it assumes a different, power law grain size distribution. The ratio of graphite over olivine dust can be varied by tuning the corresponding mass column densities $\Sigma_\text{gr}$ and $\Sigma_\text{ol}$. In particular, the mass ratio of the dust model incorporated in \textsc{skirt} (Sect.~\ref{sect:xraydustmodel}) is recovered for:
\begin{align}
    \Sigma_\text{gr} &= 0.27 \times \Sigma_\text{d}\notag \\
    \Sigma_\text{ol} &= 0.73 \times \Sigma_\text{d}\notag,
\end{align}
with $\Sigma_\text{d}$ the total mass column density of the dust.

\begin{figure*}[]
\centering
\includegraphics[width=\hsize]{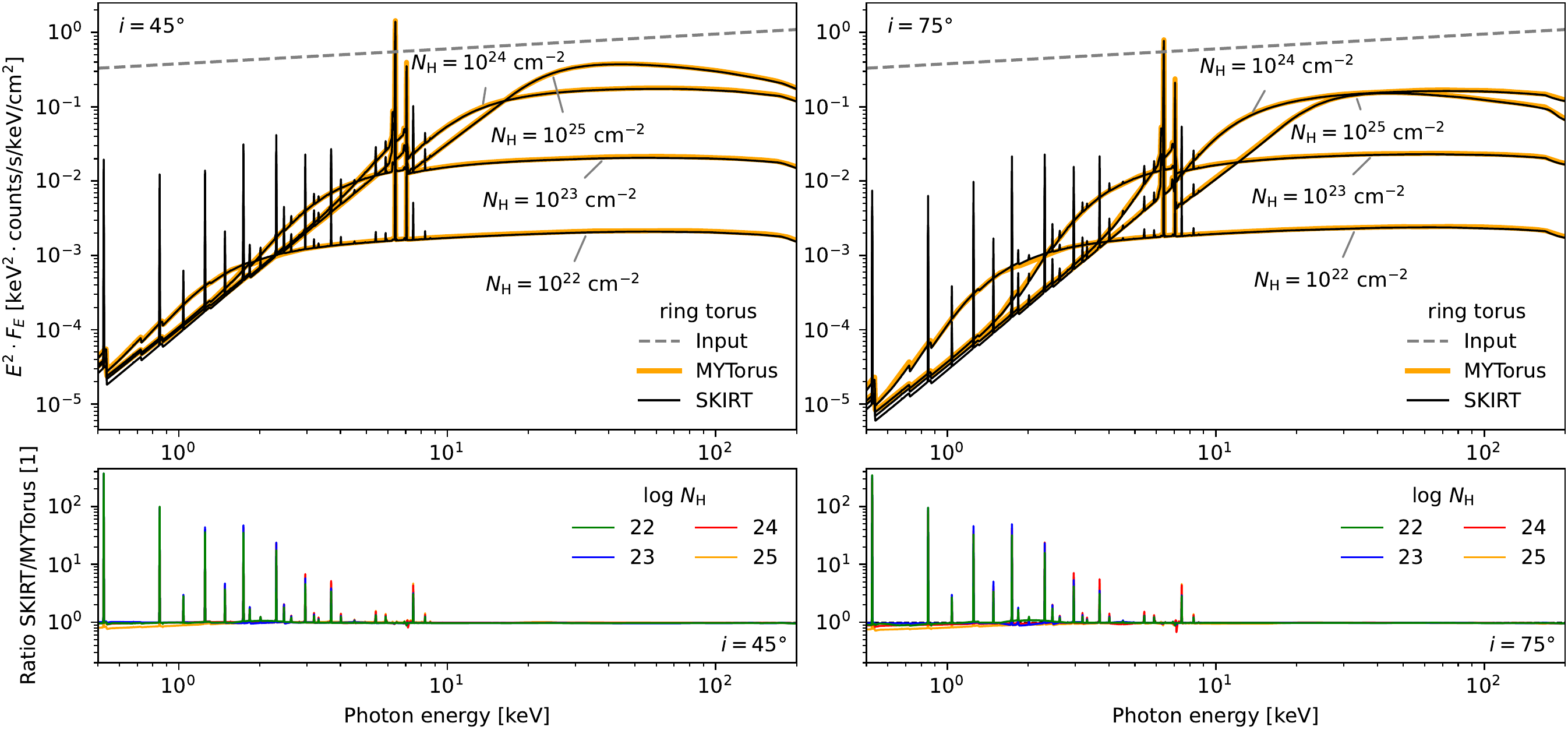}
  \caption{Fourth \textsc{skirt} benchmark (Sect.~\ref{sect:mytorus}), comparing \textsc{skirt} simulations of a torus model against the corresponding \textsc{MYTorus} model spectra. The opening angle of the ring torus is $60^\circ$. The equatorial hydrogen column density $N_\text{H}$ of the torus is indicated on the figure, for each simulation. (left) Scattered photon flux, for a line of sight not intersecting the obscuring torus ($i=45^\circ$). (right) Scattered photon flux, looking through the obscuring torus ($i=75^\circ$). We find an excellent agreement between both codes, for all column densities and sightlines.}
     \label{fig:mytorus}
\end{figure*}
\begin{figure*}[]
\centering
\includegraphics[width=\hsize]{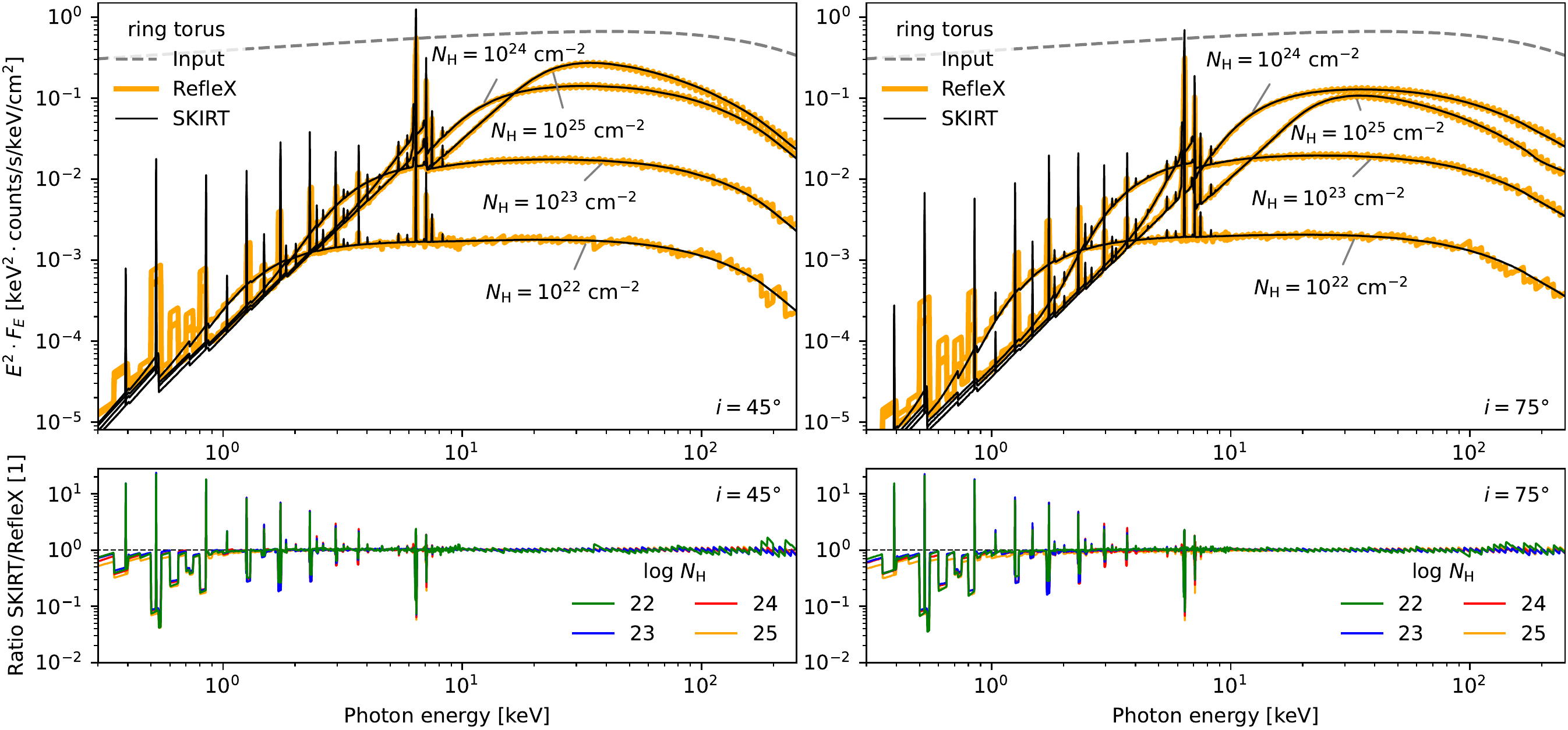}
  \caption{Fifth \textsc{skirt} benchmark (Sect.~\ref{sect:reflex}), comparing \textsc{skirt} simulations of a torus model against the corresponding \textsc{RXTorus} model spectra as calculated with \textsc{RefleX}. The opening angle of the ring torus is $60^\circ$. The equatorial hydrogen column density $N_\text{H}$ of the torus is indicated on the figure, for each simulation. (left) Scattered photon flux, for a line of sight not intersecting the obscuring torus ($i=45^\circ$). (right) Scattered photon flux, looking through the obscuring torus ($i=75^\circ$). We find an excellent agreement between both codes, for all column densities and sightlines.}
     \label{fig:reflex}
\end{figure*}
The \textsc{ismdust} model setup is mimicked in \textsc{skirt} with a three-dimensional slab of dust between the source and the observer, similar to the geometrical setup of the \textsc{phabs} and \textsc{cabs} model in Sect.~\ref{sect:phabscabs}. The corresponding \textsc{skirt} simulation results are shown in Fig.~\ref{fig:ismdust}, for three different dust mass column densities $\Sigma_\text{d}$. When comparing the direct photon fluxes, the \textsc{ismdust} model spectra are nicely recovered for photon energies above $0.29~\text{keV}$, showing detailed spectral features near the atomic absorption edges of \element[][]{C} ($0.29~\text{keV}$), \element[][]{O} ($0.54~\text{keV}$), \element[][]{Fe} ($0.72~\text{keV}$), \element[][]{Mg} ($1.31~\text{keV}$), and \element[][]{Si} ($1.85~\text{keV}$). A minor ($<8\%$) offset can be observed in the continuum flux between $0.6$ and $3~\text{keV}$, which could be attributed to the difference in grain size distribution. 

Below $0.29~\text{keV}$ (i.e.\ the \element[][]{C}~K-edge), the \textsc{skirt} and \textsc{ismdust} spectra appear to diverge. However, we note that the \textsc{ismdust} model was introduced to \textsc{xspec} for fitting observational X-ray data from the \emph{Chandra} and \emph{XMM-Newton} observatories, and was not designed to be applied below $0.3~\text{keV}$. For example, the grain size distribution does not include any small grains with sizes $<5~\text{nm}$, which would increase the extinction at the lowest X-ray energies. The \textsc{skirt} dust model on the other hand, is well-defined below $0.3~\text{keV}$, and extends down to cover the entire X-ray to mm wavelength range.

\subsection{Radiative transfer torus models}
\label{sect:torusmodels}
\subsubsection{\textsc{MYTorus}}
\label{sect:mytorus}
\textsc{MYTorus} \citep{murphy09} is an X-ray spectral model for absorption and reflection by a toroidal reprocessor of cold atomic gas, that pioneered the X-ray torus modelling field by providing the first \textsc{xspec} model for observational data fitting.\footnote{\url{http://mytorus.com/}} \textsc{MYTorus} models a ring torus (i.e.\ a doughnut) geometry of gas, with a fixed opening angle\footnote{Throughout this work, we define the torus opening angle as the half opening angle as measured from the polar axis.} of $60^{\circ}$ and a uniform density. Photo-absorption is implemented based on \citet{vern1, vern2} cross sections and \citet[][]{angr} abundances (i.e.\ similar to \textsc{skirt}, see Sect.~\ref{sect:PA}), while fluorescence is limited to the iron K$_{\alpha}$ and K$_{\beta}$ lines, and bound-electron scattering is approximated as free-electron scattering (see Sect.~\ref{sect:BES}). The primary X-ray source has a power law spectrum similar to the spectrum defined in Sect.~\ref{sect:benchmark}, but extends up to $500~\text{keV}$ without exponential cut-off.

We reproduce the \textsc{MYTorus} model in \textsc{skirt} with a smooth ring torus of cold gas, centred around an X-ray point source. We ignore bound-electron scattering, and assume \citet[][]{angr} abundances. We run \textsc{skirt} simulations for two viewing angles, so that one line of sight sees the central source unobscured ($i = 45^{\circ}$), while the other one intersects the torus ($i = 75^{\circ}$). We vary the equatorial hydrogen column density of the torus between $10^{22}$ and $10^{25}~\text{cm}^{-2}$, and use the \textsc{MYTorus} power law input spectrum as described above.

Fig.~\ref{fig:mytorus} shows the \textsc{skirt} simulation results together with the \textsc{MYTorus} reference spectra. For this comparison, we focus on the reprocessed emission only, which is more informative than the total observed spectra that include direct emission (which can be obtained analytically,\footnote{With $N_\text{H, los} = N_\text{H} \times \sqrt{1-\cos^2{i}/\cos^2{60^\circ}}$ for sightlines passing through the obscuring torus ($i>60^\circ$).} without any non-linear radiative transfer effects). We find an excellent agreement between both codes over the entire simulated energy range, for both sightlines and all torus densities. The common \element[][]{Fe} K$_{\alpha}$ and K$_{\beta}$ lines are matched perfectly, as well as the \element[][]{Fe} K$_{\alpha}$ Compton shoulder, while \textsc{skirt} implements a large number of additional fluorescent lines. The \textsc{skirt} results also reproduce the \enquote{knee} feature in the \textsc{MYTorus} spectra at $169~\text{keV}$, which is caused by the sudden lack of input photons beyond $500~\text{keV}$ to seed the Compton down-scattered continuum above $169~\text{keV}$.

For high column densities, one can observe a small offset between the \textsc{skirt} and \textsc{MYTorus} results at low photon energies. However, one should note that column densities of $10^{25}~\text{cm}^{-2}$ correspond to optical depths as high as $\tau>100$ and $\tau>2000$ for photon energies below $3$ and $1~\text{keV}$, respectively (see Fig.~\ref{fig:totalextinction}). Therefore, one cannot expect realistic soft X-ray model spectra for these heavily obscured sightlines \citep[see e.g.][]{camps18}. Finally, some differences should be expected as the \textsc{skirt} spectra are calculated for an observer at the exact inclination of $i = 45^{\circ}$ and $i = 75^{\circ}$, while the \textsc{MYTorus} model spectra where obtained by averaging the observed radiation over large angular bins centred on $12$ representative inclination values, which then need to be interpolated by \textsc{xspec}.

\subsubsection{\textsc{RefleX}}
\label{sect:reflex}
\textsc{RefleX} \citep{paltani17} is a well-established ray-tracing code, focusing on X-ray radiative transfer in the circumnuclear media of AGN. The code implements a complete set of X-ray physics in cold-gas media, including bound-electron scattering and a large collection of fluorescent line transitions. In particular, the \textsc{RefleX} code implements the same interaction cross sections as those that are incorporated in \textsc{skirt}, forming an ideal reference for benchmarking the new X-ray processes.

\textsc{RXTorus} is a smooth torus model that was calculated with \textsc{RefleX}, and is publicly available for spectral fitting.\footnote{\url{https://www.astro.unige.ch/reflex/}} The \textsc{RXTorus} model represents a ring torus of cold gas, with a uniform density and a variable covering factor, which models scattering on bound electrons (Sect.~\ref{sect:BES}), in addition to photo-absorption (Sect.~\ref{sect:PA}) and fluorescence (Sect.~\ref{sect:FL}). For a torus opening angle of $60^{\circ}$ (corresponding to an axis ratio of $r/R=0.5$), the \textsc{RXTorus} geometry is identical to the \textsc{mytorus} model geometry (Sect.~\ref{sect:mytorus}). \textsc{RXTorus} has been tested against preceding torus models such as \textsc{mytorus} \citep{murphy09}, \textsc{BNTorus} \citep{brightman11}, and \textsc{CTorus} \citep{liu14, liu15}, and was applied to the observational X-ray data of NGC~424 \citep[][]{paltani17}.

We reproduce the \textsc{RXTorus} model in \textsc{skirt} with a uniform ring torus of cold gas, centred around an X-ray point source. Consistent with the \textsc{RXTorus} model, we enable bound-electron scattering, and assume \citet[][]{angr} abundances. We fix the torus opening angle to $60^{\circ}$, and run \textsc{skirt} simulations for an unobscured ($i = 45^{\circ}$) and an obscured ($i = 75^{\circ}$) sightline. The equatorial hydrogen column density of the torus is varied between $10^{22}$ and $10^{25}~\text{cm}^{-2}$, which are the $N_\text{H}$-limits of the \textsc{RXTorus} model.

The \textsc{skirt} simulation results are shown in Fig.~\ref{fig:reflex}, together with the \textsc{RefleX} reference spectra. Both models are found to be in excellent agreement over the entire simulation domain, for both sightlines and all column densities between $10^{22}$ and $10^{25}~\text{cm}^{-2}$. The main observed difference is the high level of photon counting noise in the \textsc{RXTorus} spectra, especially when the reflected flux is low, e.g.\ when photo-absorption is high ($E < 1~\text{keV}$) or when there is little reflecting material ($N_\text{H} = 10^{22}~\text{cm}^{-2}$). The Poisson noise in the corresponding \textsc{skirt} results is significantly lower, partly due to the MCRT optimisation mechanisms that were described in Sect.~\ref{sect:MCRT}. Furthermore, the simulated \textsc{skirt} spectra have a higher spectral resolution, producing more narrow fluorescent lines and smoother reflection continua. However, both the fluorescent line fluxes and the continuum levels are found to be consistent in both radiative transfer calculations.

\begin{figure}
\centering
\includegraphics[width=\hsize]{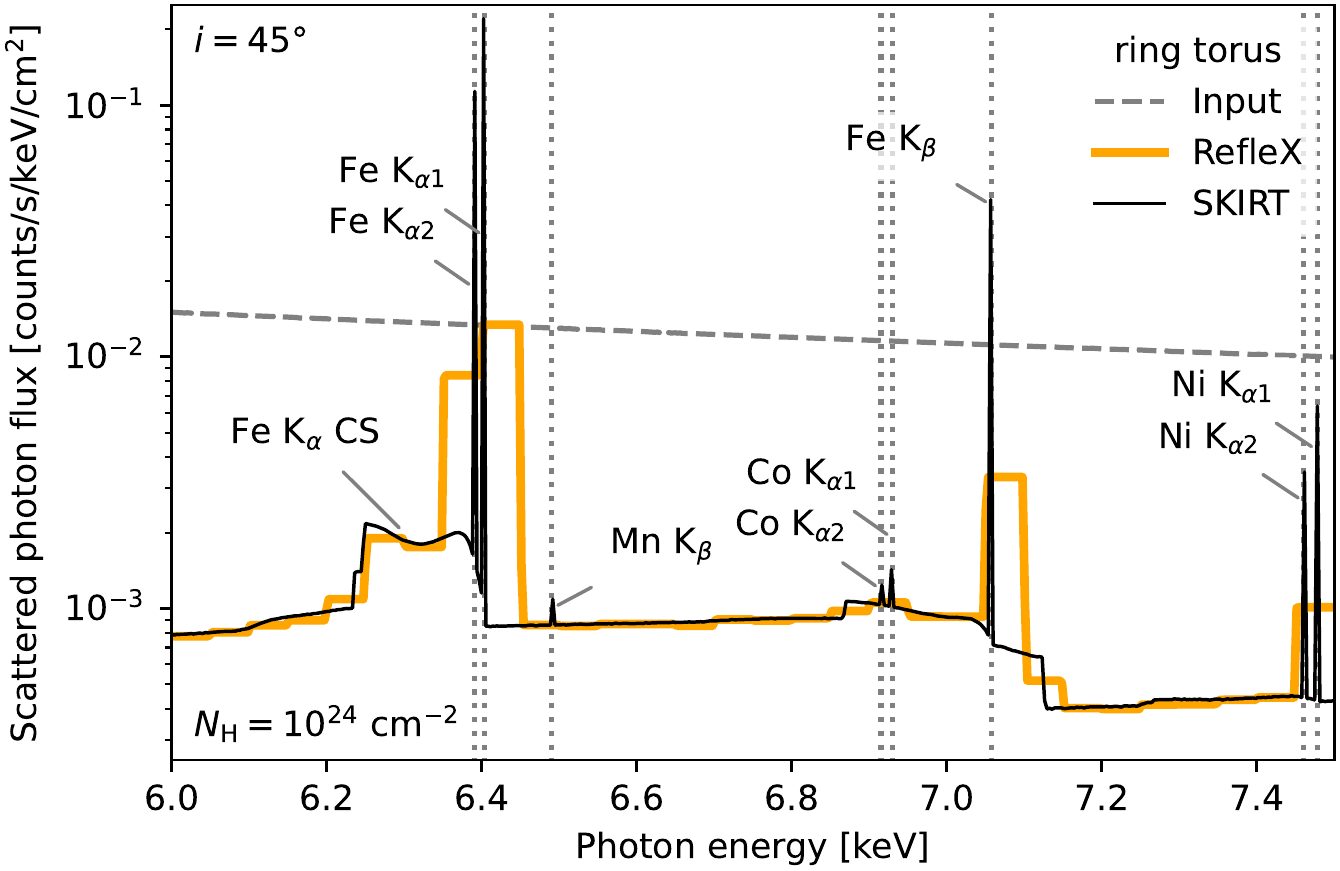}
  \caption{Zoom on the $6.0$ to $7.5~\text{keV}$ reflection spectrum of a ring torus model with $N_\text{H} = 10^{24}~\text{cm}^{-2}$ (same model as Fig.~\ref{fig:reflex}, left). Table~\ref{table:2} lists the integrated line fluxes for the three most prominent fluorescent lines. Note how \textsc{skirt} can spectrally resolve the K$_{\alpha1}$ and K$_{\alpha2}$ sublines.}
     \label{fig:reflex_CS}
\end{figure}
Fig.~\ref{fig:reflex_CS} shows a zoom into the important $6.0$ to $7.5~\text{keV}$ spectral region, for the same ring torus model with $N_\text{H} = 10^{24}~\text{cm}^{-2}$ and $i=45^\circ$. This range contains the three most prominent fluorescent lines (\element[][]{Fe} K$_{\alpha}$, \element[][]{Fe} K$_{\beta}$, and \element[][]{Ni} K$_{\alpha}$), plus the \element[][]{Fe} K$_{\alpha}$ Compton shoulder, which forms a powerful probe on the circumnuclear material, being sensitive to both the medium geometry and the physical state of the scattering electrons (bound or free). Despite the limited spectral resolution of the \textsc{RefleX} simulations, we distinguish clear Compton shoulders in both simulations, with similar strengths and consistent spectral shapes. The \textsc{skirt} spectrum reveals some additional substructure in the Compton shoulder shape, which could be observed with upcoming microcalorimeter observatories. Due to the difference in spectral resolution between \textsc{skirt} and \textsc{RefleX}, it is difficult to compare the fluorescent line strengths by eye. Therefore, we measure the integrated line fluxes for \element[][]{Fe} K$_{\alpha}$, \element[][]{Fe} K$_{\beta}$, and \element[][]{Ni} K$_{\alpha}$, which we have listed in Table~\ref{table:2}. We find an excellent agreement between the line fluxes in both codes, with relative differences in the iron line fluxes as low as $0.9\%$ (\element[][]{Fe} K$_{\alpha}$) and $0.3\%$ (\element[][]{Fe} K$_{\beta}$), while the much fainter \element[][]{Ni} K$_{\alpha}$ line deviates only $4.5\%$ in flux. Finally, we note how the superior spectral resolution of the \textsc{skirt} spectra reveals the K$_{\alpha1}$ and K$_{\alpha2}$ sublines, as shown in Fig.~\ref{fig:reflex_CS}.

\begin{table}\caption{Integrated line fluxes as calculated with \textsc{skirt} and \textsc{RefleX}, for three important fluorescent lines in the $6.0$ to $7.5~\text{keV}$ range, see Fig.~\ref{fig:reflex_CS}. The line fluxes in both radiative transfer calculations are found to be in excellent agreement.}
\label{table:2}
\centering                                     
\begin{tabular}{c c c c}          
\hline\hline                       
Line & $F_{\text{line}}$ with \textsc{skirt} $[10^{-4}]$& $F_{\text{line}}$ with \textsc{RefleX} $[10^{-4}]$\\    
\hline                                  
    \element[][]{Fe} K$_{\alpha}$ & $9.46~\text{counts}~\text{s}^{-1}~\text{cm}^{-2}$ & $9.55~\text{counts}~\text{s}^{-1}~\text{cm}^{-2}$\\     
    \element[][]{Fe} K$_{\beta}$ & $1.30~\text{counts}~\text{s}^{-1}~\text{cm}^{-2}$ & $1.30~\text{counts}~\text{s}^{-1}~\text{cm}^{-2}$\\ 
    \element[][]{Ni} K$_{\alpha}$ & $0.30~\text{counts}~\text{s}^{-1}~\text{cm}^{-2}$ & $0.29~\text{counts}~\text{s}^{-1}~\text{cm}^{-2}$\\ 
\hline                                             
\end{tabular}
\end{table}
\begin{figure*}
\centering
\includegraphics[width=\hsize]{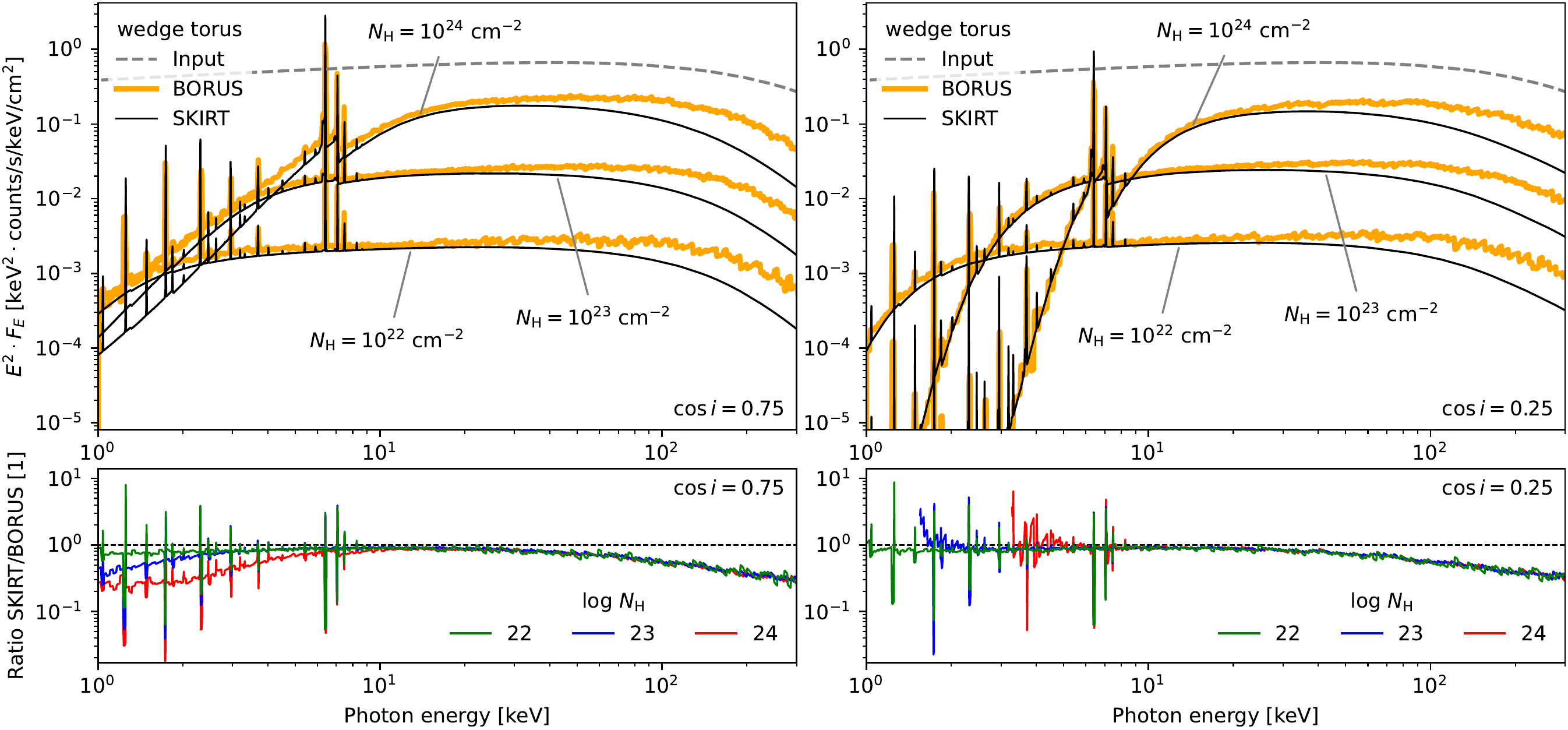}
  \caption{Sixth \textsc{skirt} benchmark (Sect.~\ref{sect:borus}), comparing \textsc{skirt} simulations of a torus model against the corresponding \textsc{borus02} model spectra as calculated with \textsc{borus}. The opening angle of the wedge torus is $60^\circ$. The equatorial hydrogen column density $N_\text{H}$ of the torus is indicated on the figure, for each simulation. (left) Scattered photon flux, for a line of sight not intersecting the obscuring torus ($\cos i=0.75$). (right) Scattered photon flux, looking through the obscuring torus ($\cos i=0.25$). Above $30~\text{keV}$, we find a significant mismatch between both codes for all sightlines. For unobscured sightlines ($\cos i=0.75$), the simulation results also diverge for energies below $10~\text{keV}$.}
     \label{fig:borus}
\end{figure*} 
\subsubsection{\textsc{borus}}
\label{sect:borus}
\textsc{borus} \citep{balokovic18, balokovic19} is a recent suite of X-ray spectral models for fitting observational data of obscured AGN. Their most popular \textsc{borus02} model implements a uniform-density sphere with conical cutouts at the poles, forming a wedge torus (i.e.\ a flared disk geometry). This torus geometry is identical to the \textsc{BNTorus} \citep{brightman11} model, and is publicly available as an \textsc{xspec} table model.\footnote{\url{https://sites.astro.caltech.edu/~mislavb/download/}} \textsc{borus02} has similar model parameters as the \textsc{RXTorus} model described in Sect.~\ref{sect:reflex}, with additional control over the exponential cut-off energy and the \element[][]{Fe} abundance. However, \textsc{borus02} is more restricted in terms of X-ray physics, as it ignores bound-electron scattering, and implements the \textsc{nist/xcom}\footnote{\url{https://physics.nist.gov/PhysRefData/Xcom/html/xcom1.html}} photo-absorption cross sections that are limited to $E\ge1~\text{keV}$. The \textsc{borus02} torus geometry forms the most advanced transfer geometry within \textsc{borus}, with one more spectral model representing a uniform-density sphere. 
 
We reproduce the \textsc{borus02}\footnote{We used the most recent \textsc{borus02} tables (version 170323a).} model in \textsc{skirt} with a uniform wedge torus of cold gas, assuming \citet[][]{angr} gas abundances and free-electron scattering. We fix the torus opening angle to $60^{\circ}$, corresponding to a covering factor of $50\%$. We run \textsc{skirt} simulations for unobscured ($\cos i=0.75$) and obscured ($\cos i=0.25$) sightlines, and vary the equatorial hydrogen column density of the torus between $10^{22}$ and $10^{24}~\text{cm}^{-2}$. These parameter combinations are grid points of the \textsc{borus02} table model, for which the \textsc{borus} radiative transfer code was previously run. In this way, we assure that our benchmark results are unaffected by interpolation effects between gridded \textsc{borus02} realisations. Fig.~\ref{fig:borus} shows the \textsc{skirt} simulation results and the \textsc{borus} reference spectra. The \textsc{borus02} model provides scattered photon fluxes without the direct emission component, which allows for a direct comparison of the X-ray reflection spectra as calculated with \textsc{skirt} and \textsc{borus}.

We find significant discrepancies between the results obtained with \textsc{skirt} and \textsc{borus}, that are not really understood (see Fig.~\ref{fig:borus}). First, we observe a mismatch between the \textsc{skirt} and \textsc{borus} reflection spectra for $E>30~\text{keV}$, which is present for all torus densities and both viewing angles. In this energy range, the reflected X-ray emission is governed by Compton scattering, with photo-absorption being negligible (Fig.~\ref{fig:totalextinction}). Yet, we find the Compton-reflected \textsc{skirt} continua to be $20$ to $70\%$ lower than the corresponding \textsc{borus} results in the $30$ to $300~\text{keV}$ range. These high energy differences where traced back to a \textsc{borus02} error in the convolution of the \textsc{borus} Green functions with the input spectrum, which will be corrected in an upcoming \textsc{borus02} table release (Balokovi{\'c}, priv. comm.).

Furthermore, we find that the \textsc{skirt} fluxes at lower energies ($E<10~\text{keV}$) are also disagreeing with the \textsc{borus} results, although for unobscured sightlines only ($\cos i=0.75$, as shown in the left panels of Fig.~\ref{fig:borus}). In this energy range, the reflected X-ray continuum is produced by Compton scattering, but is then heavily photo-absorbed by the intervening gas. Remarkably, we do find a good agreement between both codes for obscured sightlines in the same energy range (Fig.~\ref{fig:borus}, right). Finally, we obtain an excellent agreement over the $6.0$ to $7.5~\text{keV}$ spectral range for obscured sightlines, reproducing the characteristic spectral shape of heavily obscured AGN with both codes (Fig.~\ref{fig:borus}, right).

Our simulated \textsc{skirt} spectra have a much higher spectral resolution than the publicly-available \textsc{borus02} tables, as needed to model the highest resolution CCD-based spectra and upcoming microcalorimeter observations with \emph{XRISM}. Even with this larger number of spectral bins, however, our \textsc{skirt} results still exhibit less numerical noise.

\subsubsection{\textsc{xars}}
\label{sect:xars}
Most recently, the Python-based \textsc{xars} code \citep{buchner19} introduced several new medium geometries to X-ray radiative transfer simulations. \textsc{xars} is a Monte Carlo radiative transfer code designed to study obscuration variability in AGN, which applies the same simulation method as \textsc{BNTorus} \citep{brightman11}. The \textsc{xars} code can handle non-axisymmetric transfer media such as warped disks, clumpy tori, and arbitrary constellations of spherical clumps, and was used to compare hydrodynamical models against observational data of Circinus galaxy \citep{buchner21}. The \textsc{xars} code is open-source and publicly available online\footnote{\url{https://github.com/JohannesBuchner/xars}}, and can easily be extended with new transfer geometries. 

 On the other hand, this Monte Carlo code does not implement some of the well-known biasing techniques that make MCRT simulations efficient in 3D. For example, \textsc{xars} allows photons to escape the model in all directions, and then averages their contributions over direction bins to obtain the observed flux. Such a detection scheme is inefficient compared to the photon peel-off method (see Sect.~\ref{sect:MCRT}), especially for non-axisymmetric transfer media. Compared to the \textsc{RefleX} code (Sect.~\ref{sect:reflex}), \textsc{xars} does not implement bound-electron scattering, while fluorescence is limited to 11 line transitions, based on fluorescent yields by \citet[][]{bambynek72}.

\begin{figure*}
\centering
\includegraphics[width=\hsize]{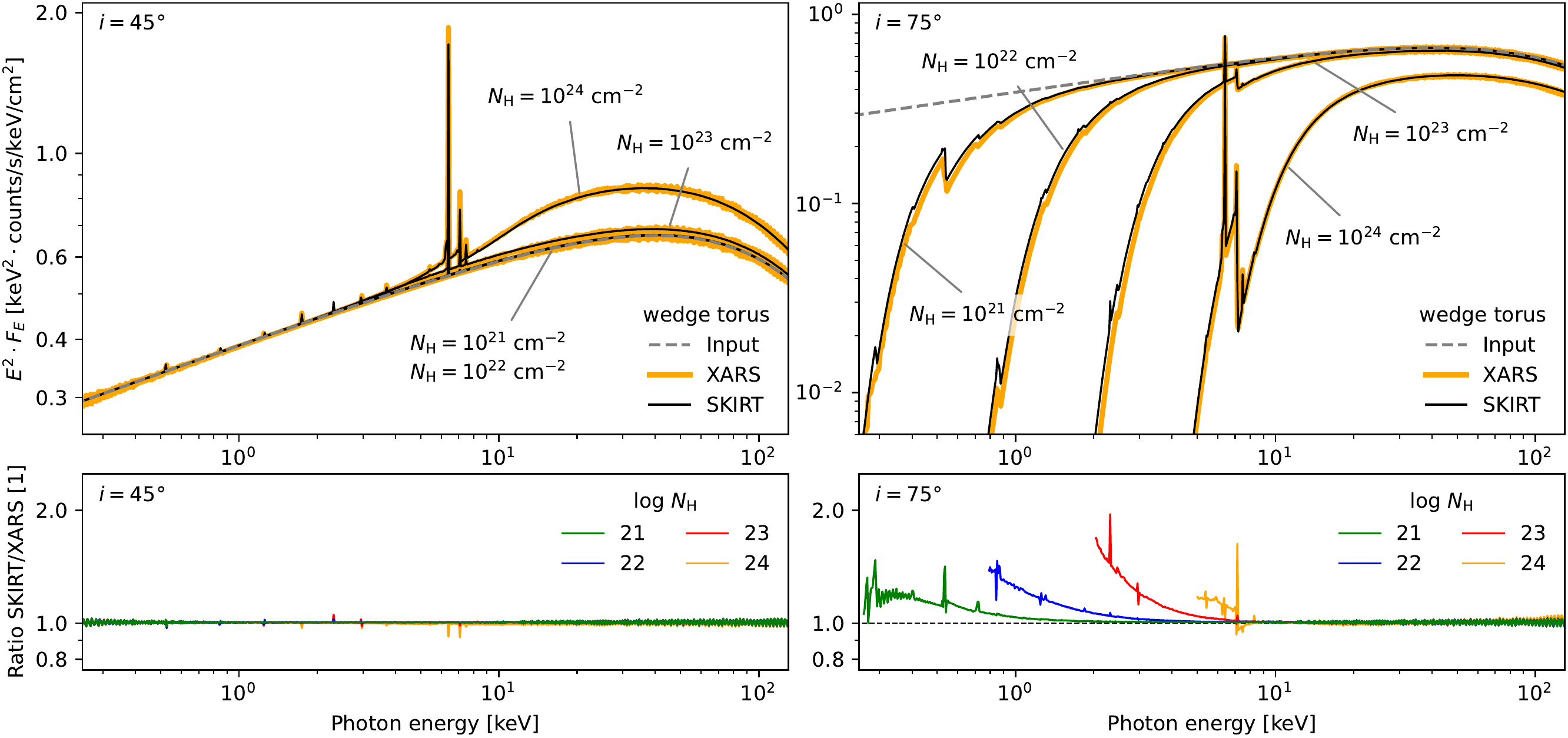}
  \caption{Seventh \textsc{skirt} benchmark (Sect.~\ref{sect:xars}), comparing \textsc{skirt} simulations of a torus model against the corresponding \textsc{xars} spectra. The opening angle of the wedge torus is $60^\circ$. The equatorial hydrogen column density $N_\text{H}$ of the torus is indicated on the figure, for each simulation. (left) Total photon flux, for a line of sight not intersecting the obscuring torus ($i=45^\circ$). (right) Total photon flux, looking through the obscuring torus ($i=75^\circ$). Both codes are in excellent agreement. The small offset below $12.5~\text{keV}$ is explained by differences in the atomic data (see text).}
     \label{fig:xars}
\end{figure*}

We compare \textsc{skirt} against the \textsc{xars} code, focusing on the \textsc{wedge-cutoff} torus model, which represents a uniform torus with a variable opening angle similar to the \textsc{borus02} model (see Sect.~\ref{sect:borus}). We reproduce the model in \textsc{skirt} with a uniform wedge torus of cold gas with an opening angle of $60^\circ$, identical to the geometrical setup in Sect.~\ref{sect:borus}. We assume free-electron scattering and \citet[][]{angr} gas abundances, and vary the equatorial hydrogen column density between $10^{21}$ and $10^{24}~\text{cm}^{-2}$. The \textsc{skirt} simulation results are shown in Fig.~\ref{fig:xars}, together with the \textsc{xars} reference spectra for two different viewing angles ($45^{\circ}$ and $75^{\circ}$). As the \textsc{wedge-cutoff} model does not provide the scattered flux as a separate model component, we compare the total observed fluxes as calculated with both codes. Note how this causes the reflection spectrum to be hidden by the strong direct component for $N_\text{H} < 10^{23}~\text{cm}^{-2}$. 

We find a good agreement between the \textsc{skirt} and \textsc{xars} results, for both sightlines and all torus densities. The main observed difference is the slightly higher degree of photo-absorption in the \textsc{xars} spectra at soft X-ray energies, as shown in the right panels of Fig.~\ref{fig:xars}. This offset can be traced back to the photo-absorption cross sections implemented in \textsc{xars}, which differ from the \textsc{phfit2}\footnote{\textsc{phfit2} is the original Fortran routine implementing the cross sections of \citet[][]{vern1} and \citet{vern2}, see Sect.~\ref{sect:PA}.} cross sections used in \textsc{skirt}, \textsc{RefleX}, and \textsc{xspec}. The \textsc{xars} cross sections are $10$ to $15\%$ higher for photon energies below $12.5~\text{keV}$, causing the corresponding soft X-ray fluxes to be lower for obscured sightlines. The induced flux reduction is expected to further depend on the column density and the photon energy, as shown in the bottom right panel of Fig.~\ref{fig:xars}. 

At higher photon energies, the total gas extinction is fully dominated by Compton scattering (see Fig.~\ref{fig:bes} and Fig.~\ref{fig:totalextinction}), which is treated consistently in both codes and therefore reconciles the \textsc{skirt} and \textsc{xars} results. In particular, the Compton reflection hump peaking at about $40~\text{keV}$ is nicely recovered on top of the unobscured input spectrum with both codes, as shown in the left top panel of Fig.~\ref{fig:xars}. This same figure shows a clear \element[ ][]{S}~K$_\alpha$ line in the \textsc{skirt} results at $2.3~\text{keV}$, which is the strongest fluorescent line that is not implemented in \textsc{xars}, arguably more important than the available \element[][]{Ne} and \element[][]{Mg} transitions. Finally, we find an excellent agreement between both codes in the $6.0$ to $7.5~\text{keV}$ spectral region for heavily obscured sightlines, reproducing the characteristic spectral shape of heavily obscured AGN (top right panel of Fig.~\ref{fig:xars}, for $N_\text{H} = 10^{24}~\text{cm}^{-2}$). Overall, less Monte Carlo noise is observed in the \textsc{skirt} results, which can be linked to the implemented optimisation techniques (Sect.~\ref{sect:MCRT}).

\section{Demonstration}
\label{sect:demonstration}
We illustrate the 3D capabilities of \textsc{skirt} in the X-ray regime by simulating a set of clumpy torus models, where a variable fraction of the torus material is concentrated in clumps. These clumps are randomly distributed over the torus, forming a truly three-dimensional transfer medium with no spatial symmetries to reduce the dimensionality of the radiative transfer simulations. 

The baseline smooth torus model is a uniform-density wedge torus of cold gas with an opening angle of $60^\circ$, and an equatorial hydrogen column density of $5 \times 10^{22}~\text{cm}^{-2}$, which is centred around an X-ray point source with $\Gamma=1.8$, $E_\textrm{cut}=200~\text{keV}$, and $L_{2-10}=2.8 \times 10^{42}~\text{erg}~\text{s}^{-1}$ (see Sect.~\ref{sect:benchmark}). The radial extent of the torus is $15~\text{pc}$, and the gas abundances are assumed to be solar, with bound-electron scattering enabled. We then consider the corresponding two-phase torus model by allocating $50\%$ of the torus mass to $1000$ spherical clumps of $0.5~\text{pc}$ radius, and the clumpy torus model that has all its mass in clumps \citep[see also][]{stalevski12}. Within each clump, we assume a cubic spline radial density profile. For details on the implementation of such a clumpy geometry we refer to \citet{baes15}.

\begin{figure*}
\centering
\includegraphics[width=\hsize]{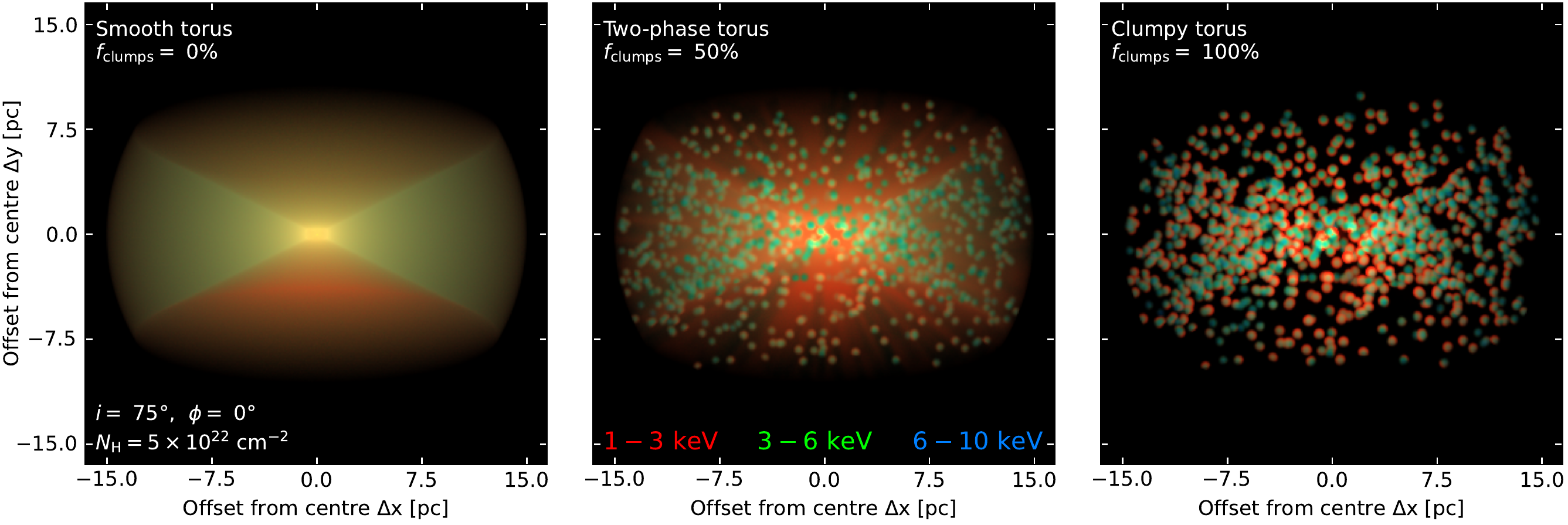}
  \caption{\textsc{skirt} X-ray images of a smooth torus model (left), two-phase torus model (middle), and clumpy torus model (right), observed at $i=75^\circ$. The RGB colours correspond to the integrated fluxes in the given ranges. Clumpy AGN tori represent truly three-dimensional transfer geometries with no spatial symmetries.}
     \label{fig:demo}
\end{figure*}
Fig.~\ref{fig:demo} shows the \textsc{skirt} X-ray images for the smooth, two-phase, and clumpy torus models described above. The smooth torus model (Fig.~\ref{fig:demo}, left) exhibits a colour asymmetry between the illuminated back and front side of the torus, which directly relates to the X-ray scattering physics in the gas (Sect.~\ref{sect:BES}). The two-phase and clumpy torus models show more complex morphologies, and demonstrate how individual clumps can be properly resolved in 3D radiative transfer simulations with \textsc{skirt}. These clumps form dense scattering sites inside the torus, which are not easily penetrated by soft X-ray photons, producing radially extended shadow lanes (Fig.~\ref{fig:demo}, middle). For the two-phase and clumpy torus models, the column densities of individual clumps reach $9.0 \times 10^{22}~\text{cm}^{-2}$ and $1.8 \times 10^{23}~\text{cm}^{-2}$, respectively. Overall, these images illustrate how the entire torus structure contributes to the X-ray reflection spectrum, through multiple scattering inside the three-dimensional transfer medium.

\begin{figure}
\centering
\includegraphics[width=\hsize]{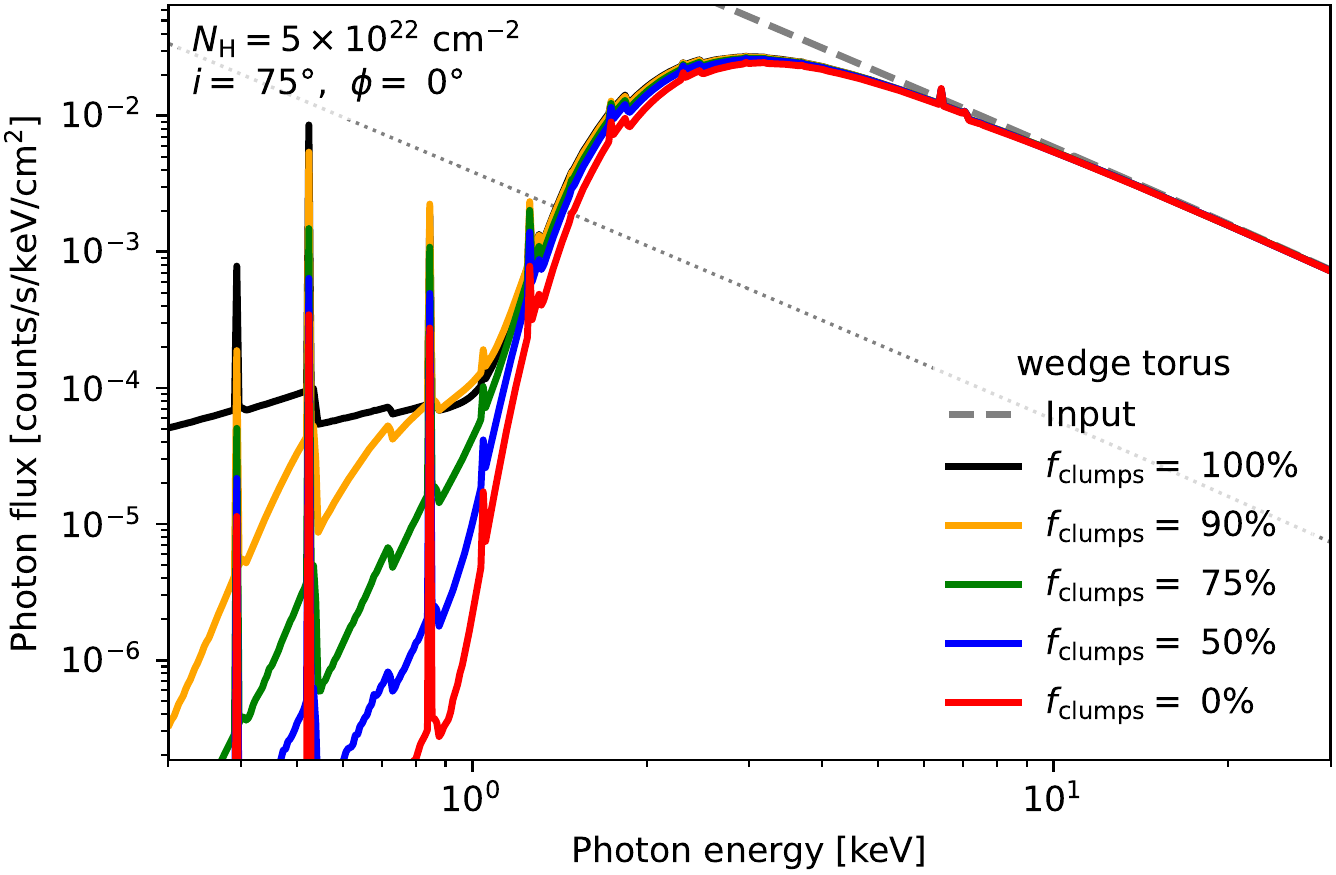}
  \caption{\textsc{skirt} X-ray spectra for torus models with different clump mass fractions $f_\mathrm{clumps}$. The red, blue, and black spectra correspond to the smooth, two-phase, and clumpy tori in Fig.~\ref{fig:demo}, respectively. The grey dotted line represents additional Thomson scattered emission at $1\%$ of the intrinsic continuum \citep[see][]{gupta21}.}
     \label{fig:ClumpyFclumps}
\end{figure}
\begin{figure}
\centering
\includegraphics[width=\hsize]{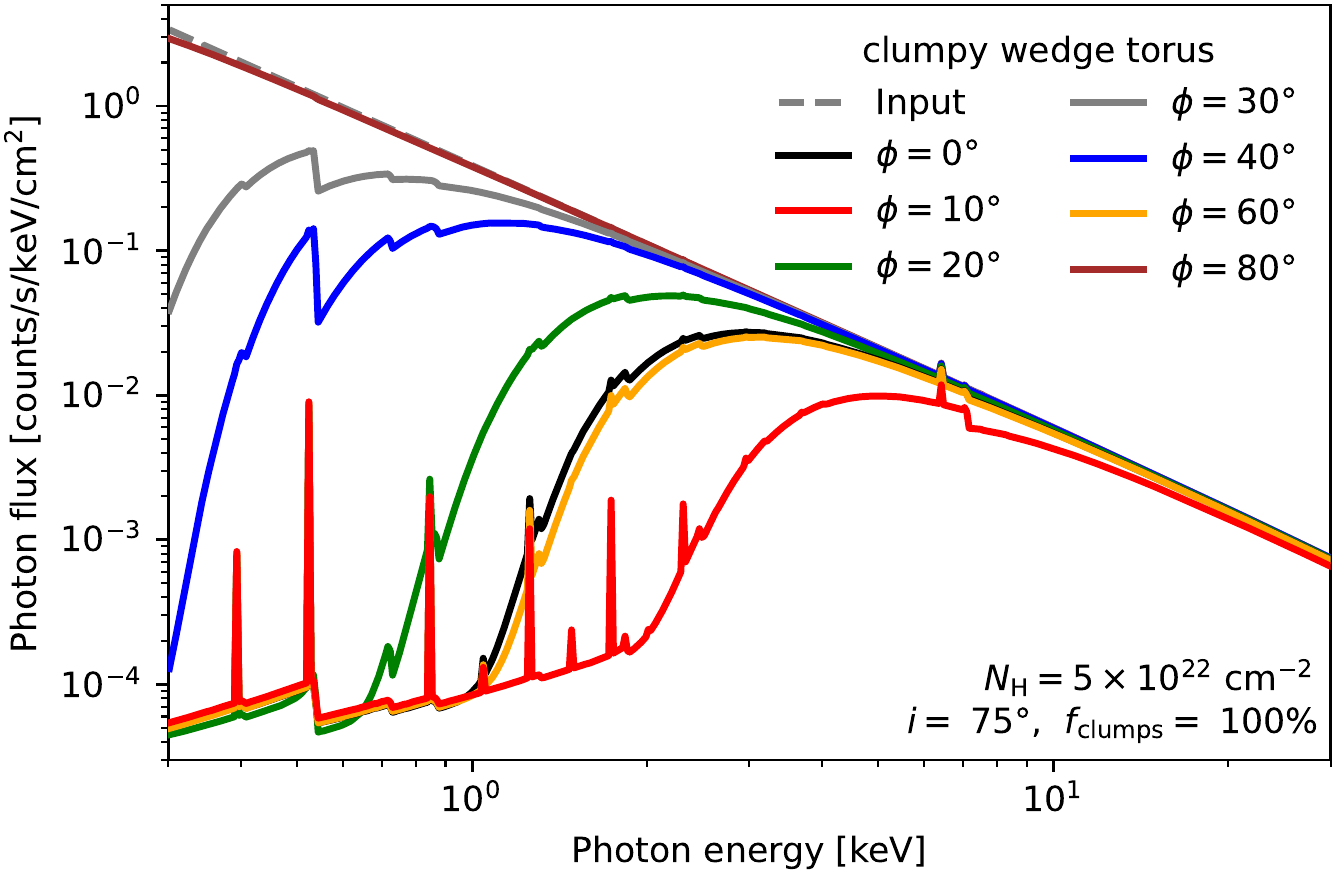}
  \caption{\textsc{skirt} X-ray spectra for a clumpy torus model ($f_\mathrm{clumps}=100\%$), observed from different azimuthal viewing angles (with $i=75^\circ$ fixed).}
     \label{fig:ClumpyAzimuths}
\end{figure} 
The corresponding X-ray spectra are shown in Fig.~\ref{fig:ClumpyFclumps}, demonstrating how clumpy torus models allow some soft X-ray photons to escape through the less dense inter-clump material. Note how the simulated differences between these spectra might be difficult to distinguish in real observations, due to the presence of additional strong components in the soft X-ray range, such as Thomson scattered emission originating from larger distances \citep[see][and Fig.~\ref{fig:ClumpyFclumps}]{gupta21}. Finally, we illustrate how torus models lose their axial symmetry when making them clumpy, by observing one clumpy torus ($f_\mathrm{clumps}=100\%$) from different azimuthal viewing angles in Fig.~\ref{fig:ClumpyAzimuths}. Depending on the distribution of clumps along the line of sight to the observer, the observed X-ray spectra can differ drastically. For example, the line of sight corresponding to $\phi = 80^\circ$ does not intersect any clumps, whereas the one corresponding to $\phi=10^\circ$ intersects several clumps, which leads to an attenuation of the soft X-ray flux at $1~\text{keV}$ by more than $3$ orders of magnitude. We refer to future \textsc{skirt} work for a more in-depth systematic study of different 3D torus geometries, under various physical assumptions. 

\section{Discussion and conclusions}
\label{sect:discussion}
We presented the new X-ray functionalities of the MCRT code \textsc{skirt}, for studying cold dusty gas in complex three-dimensional transfer media. Focusing on continuum radiative transfer in the $0.1$ to $500~\text{keV}$ range, we introduced Compton scattering on free electrons (Sect.~\ref{sect:CS}), photo-absorption (Sect.~\ref{sect:PA}) and fluorescence (Sect.~\ref{sect:FL}) by cold atomic gas, scattering on bound electrons (Sect.~\ref{sect:BES}), and extinction by dust (Sect.~\ref{sect:xraydustmodel}). This includes a novel treatment of extreme-forward scattering by dust (Sect.~\ref{sect:extremeforwardscattering}), and a detailed description of anomalous Rayleigh scattering (Sect.~\ref{sect:BES}). By extending the \textsc{skirt} code with these X-ray processes, we obtained an X-ray radiative transfer code with all features of the established \textsc{skirt} framework (Sect.~\ref{sect:MCRT}), fully optimised to operate in general 3D geometries. Furthermore, by extending the wavelength domain of \textsc{skirt} into the X-ray range, we enabled radiative transfer simulations that can self-consistently cover the X-ray band plus the IR-to-UV wavelength range, linking X-ray reprocessing to dust modelling.

The main motivation for extending the advanced treatment of absorption, scattering, and re-emission into the X-ray range, was to study the structure of AGN circumnuclear media based on their reprocessed X-ray emission. With \textsc{skirt}, we can now experiment with X-ray obscuration models of arbitrary complexity, to challenge the classical torus paradigm in a computation-efficient way. Indeed, the geometrical flexibility of \textsc{skirt} allows for testing a broad suite of spectral models featuring clumps and filaments, polar structures, and distributions imported from hydrodynamical simulations. Next to this 3D-efficiency, \textsc{skirt} offers a complete set of X-ray physics for modelling gas and dust in AGN circumnuclear media, exceeding what is currently implemented in most X-ray spectral models. Apart from AGN, we foresee additional applications for the \textsc{skirt} code in the X-ray band, e.g.\ in modelling galactic sources such as novae and X-ray binaries, or modelling the X-ray SEDs of galaxies. Finally, the advanced treatment of extreme-forward scattering by dust will make \textsc{skirt} a powerful tool for studying X-ray scattering halos formed in complex dust geometries.

We verified our X-ray implementation by reproducing some well-established \textsc{xspec} models, and found that our \textsc{skirt} results were perfectly consistent with the standard spectral models for free-electron extinction (\textsc{cabs}, Sect.~\ref{sect:phabscabs}), cold-gas absorption (\textsc{phabs}, Sect.~\ref{sect:phabscabs}), fluorescence (\textsc{pexmon}, Sect.~\ref{sect:pexmon}), Compton scattering (\textsc{pexmon}, Sect.~\ref{sect:pexmon}), and dust extinction (\textsc{ismdust}, Sect.~\ref{sect:ismdust}). These one-dimensional \textsc{xspec} models incorporate few non-trivial radiative transfer effects, but assure a correct implementation of the basic X-ray processes in free-electron, cold-gas, and dust media.

We performed the first dedicated benchmark of X-ray torus models, comparing five popular X-ray radiative transfer codes (\textsc{MYTorus}, \textsc{RefleX}, \textsc{borus}, \textsc{xars}, and \textsc{skirt}). We found an excellent agreement between the \textsc{skirt} results and the \textsc{MYTorus} (Sect.~\ref{sect:mytorus}) and \textsc{RefleX} (Sect.~\ref{sect:reflex}) model spectra, for various parameter combinations. \textsc{RefleX} implements bound-electron scattering just as \textsc{skirt} and uses the same atomic data, which makes it ideal for benchmarking the new X-ray processes in \textsc{skirt}. We recovered both the reflected X-ray continuum (Fig.~\ref{fig:reflex}), the fluorescent line fluxes (Table~\ref{table:2}), and the \element[][]{Fe} K$_{\alpha}$ Compton shoulder (Fig.~\ref{fig:reflex_CS}), forming a convincing validation of the underlying X-ray implementation.

We found a good agreement between the \textsc{skirt} and \textsc{xars} results modelling a wedge torus medium, which is one of the many transfer geometries that was released with \textsc{xars} (Sect.~\ref{sect:xars}). Compared to \textsc{skirt}, we observed slightly more photo-absorption in the \textsc{xars} spectra at soft X-ray energies, which we traced back to a difference in the adopted photo-absorption cross sections. At higher energies, we found an excellent agreement between the Compton reflection humps as obtained with both codes (Fig.~\ref{fig:xars}), reconciling both implementations of free-electron scattering. In Sect.~\ref{sect:borus}, we found significant discrepancies between the \textsc{skirt} and \textsc{borus} simulation results, which were partly understood. These differences, observed at low and high energies (Fig.~\ref{fig:borus}), plus the minor soft X-ray offset relative to \textsc{xars}, illustrate the complexity of X-ray radiative transfer and motivate the need for a robust framework that can handle non-linear radiative transfer effects in full 3D. 

Most models featured in our X-ray torus benchmark showed more numerical noise than the corresponding \textsc{skirt} results, which should be linked to the MCRT optimisations implemented in \textsc{skirt} (Sect.~\ref{sect:MCRT}). In Monte Carlo simulations, the signal-to-noise ratio is expected to scale as the square root of the total number of photon packets, while increasing the total number of photons affects the simulation time linearly. MCRT acceleration techniques allow for reaching similar noise levels with a reduced number of photons, thus accelerating the simulation process. For a fixed number of photons, this means that \textsc{skirt} can run radiative transfer simulations at a higher spectral resolution, while maintaining reasonable noise levels over more spectral bins. In this work, all \textsc{skirt} spectra were calculated at a higher spectral resolution than the corresponding reference model, and contain significantly less noise.

In addition to this MCRT-efficiency, the \textsc{skirt} code is highly optimised in a purely computational sense, resulting in shorter simulation times for a fixed number of photon packets. This reduces the computational cost of radiative transfer simulations, which is important for calculating large model libraries. In this work, each \textsc{skirt} simulation was run on a $2.3~\text{GHz}$ 8-core laptop in just a few hours, which is remarkably short considering that each simulation featured as many as $5 \times 10^9$ photons. Furthermore, these $5 \times 10^9$ photons are directly contributing to the simulation output, which should not be compared to the total number of photons escaping the model in all directions. Increasing the total number of photons linearly affects the simulation time, but does not affect the memory requirements (assuming the same host computing system). Increasing the spectral resolution has a very limited effect on the overall memory usage, but does require launching extra photons to properly sample each energy bin. Finally, \textsc{skirt} scales efficiently on multiple cores and multiple nodes, allowing to significantly increase the computational resources when needed \citep{parallel}.

MCRT simulations are mostly limited by extreme optical depths of $\tau > 20$, depending on the model geometry and the relative placement of sources and detectors \citep{camps18}. For hard X-ray photons, this corresponds to column densities of $N_\text{H} > 2 \times 10^{25}~\text{cm}^{-2}$ (see Fig.~\ref{fig:totalextinction}), which is sufficiently high to cover the entire column density distribution of AGN in the local universe \citep{ricci15}. Indeed, \textsc{skirt} can produce transmission spectra for line-of-sight column densities of $N_\text{H} = 10^{25}~\text{cm}^{-2}$ without significant numerical noise, as shown in Fig.~\ref{fig:phabscabs}. Note how these density constraints hold for the transmitted flux component only, with no similar restrictions for the reflected flux, as demonstrated in Sect.~\ref{sect:pexmon} where reflection on an dense slab of $N_\text{H} = 10^{27}~\text{cm}^{-2}$ was modelled.

In Sect.~\ref{sect:demonstration}, we demonstrated the full-3D capabilities of \textsc{skirt} in the X-ray regime by simulating X-ray images and spectra of clumpy torus models, illustrating how \textsc{skirt} will be used to build spectral models based on complex transfer geometries. All configuration elements (such as medium geometries) needed to run these clumpy torus models, plus the benchmark simulations presented in Sect.~\ref{sect:benchmark}, are built into the public master branch of the \textsc{skirt} code, and can used without any further modifications.

The aforementioned features make \textsc{skirt} a promising new tool for modelling the complex structure of AGN circumnuclear media based on their X-ray emission. With \textsc{skirt}, we can predict fluorescent lines and Compton shoulder shapes that cannot be observed with the current generation of X-ray observatories, but should appear in upcoming microcalorimeter observations \citep[][]{barret18, tashiro20}. Arguably, the access to full-3D models of complex obscuring structures will be crucial for modelling high-resolution \emph{XRISM}/Resolve data, where we expect informative spectral features that cannot be reproduced with 1D models. The presented update of the \textsc{skirt} code allows for uncomplicated access to a wide suite of 3D X-ray models for obscured AGN, that can easily be tested and modified with limited computational efforts.

\begin{acknowledgements}
      We wish to thank the anonymous referee for their careful reading and constructive comments that helped improve the presentation of this work. B.\ V.\ acknowledges support by the Fund for Scientific Research Flanders (FWO-Vlaanderen, project 11H2121N). M.\ S.\ acknowledges support by the Science Fund of the Republic of Serbia, PROMIS 6060916, BOWIE and by the Ministry of Education, Science and Technological Development of the Republic of Serbia through the contract No. 451-03-9/2022-14/200002. We wish to thank K.\ A.\ Arnaud, B.\ T.\ Draine, A.\ A.\ Zdziarski, L.\ Corrales and M.\ Balokovi{\'c} for helpful discussions. 
\end{acknowledgements}

\bibliographystyle{aa}
\bibliography{ref}

\end{document}